\documentclass{aa}
\usepackage{graphicx}
\usepackage{amssymb,amsmath}
\usepackage{natbib}
\usepackage{txfonts}
\usepackage{url}
\usepackage{pifont}
\usepackage{multirow}
\usepackage{hhline}

\def\rj{r_{\rm j}}
\def\vj{v_{\rm j}}
\def\ri{r_{\rm i}}
\def\vi{v_{\rm i}}

\def\Msun{ M_\odot}
\def\Rsun{ R_\odot}

\def\Mjup{ M_{\rm J}}
\def\Rjup{R_{\rm J}}

\def\Mp{M_{\rm p}}
\def\Rp{R_{\rm p}}

\def\Mpj{M_{{\rm p}_{\rm j}}}
\def\Rpj{R_{{\rm p}_{\rm j}}}

\def\Ipj{I_{{\rm p}_{\rm j}}}
\def\Ip{I_{{\rm p}}}

\def\mupj{\mu_{{\rm p}_{\rm j}}}

\def\kpj{k_{2,{\rm p}_{\rm j}}}
\def\kfpj{k_{2f,{\rm p}_{\rm j}}}

\def\kfp{k_{2f,{\rm p}}}
\def\kp{k_{2,{\rm p}}}

\def\ks{k_{2,\star}}
\def\kfs{k_{2f,\star}}

\def\tearth{\tau_{\oplus}}
\def\tauHJ{\tau_{\rm HJ}}
\def\taupj{\tau_{{\rm p}_{\rm j}}}
\def\taup{\tau_{{\rm p}}}
\def\taus{\tau_{\star}}

\def\pj{{\rm p}_{\rm j}}
\def\pli{{\rm p}_{\rm i}}

\def\Op{\Omega_{\rm p}}
\def\Opj{\Omega_{\pj}}

\def\oblp{\epsilon_{\rm p}}
\def\oblpj{\epsilon_{{\rm p}_{\rm j}}}

\def\Cpj{C_{{\rm p}_{\rm j}}}

\def\Ms{M_{\star}}
\def\Rs{R_{\star}}

\def\mus{\mu_{\star}}

\def\Mearth{ M_\oplus}
\def\Rearth{ R_\oplus}

\def\Os{\Omega_{\star}}
\def\Osz{\Omega_{\star,z}}

\def\searth{\sigma_{\oplus}}

\def\ss{\sigma_{\star}}
\def\sbd{\sigma_{{\rm BD}}}
\def\Tp{T_{\rm p}}

\def\G{\mathcal{G}}

\def\d{\mathrm{d}}
\def\i{\mathrm{i}}

\def\deg{^\circ}

\usepackage[normalem]{ulem}
\usepackage{color}
\definecolor{blue}{RGB}{0,0,255}
\definecolor{red}{RGB}{255,0,0}
\definecolor{green}{RGB}{0,200,0}
\definecolor{black}{RGB}{0,0,0}

\newcommand{\cmark}{\ding{51}}%
\newcommand{\xmark}{\ding{55}}%

\newcommand{\mTp}{\textit{Mercury-T}\xspace}
\newcommand{\mercp}{\textit{Mercury}\xspace}

\begin{document}

\title{\mTp: A new code to study tidally evolving multiplanet systems. Applications to Kepler-62.} 
\titlerunning{Tidally evolving multiplanet systems. Applications to Kepler-62}

   \subtitle{ }

   \author{Emeline Bolmont \inst{1,2,3}
          \and Sean N. Raymond \inst{1,2} 
          \and Jeremy Leconte \inst{4,5,6}
          \and Franck Hersant \inst{1,2} 
          \and Alexandre C. M. Correia \inst{7,8}
                }
\authorrunning{Emeline Bolmont, et al.}

\institute{Univ. Bordeaux, LAB, UMR 5804, F-33270, Floirac, France
\and CNRS, LAB, UMR 5804, F-33270, Floirac, France
\and Now at: NaXys, Department of Mathematics, University of Namur, 8 Rempart de la Vierge, 5000 Namur, Belgium
\and Canadian Institute for Theoretical Astrophysics, 60st St George Street, University of Toronto, Toronto, ON, M5S3H8, Canada
\and Banting Fellow
\and Center for Planetary Sciences, Department of Physical \& Environmental Sciences, University of Toronto Scarborough, Toronto, ON, M1C 1A4
\and CIDMA, Departamento de F\'isica, Universidade de Aveiro, Campus de Santiago, 3810-193 Aveiro, Portugal
\and ASD, IMCCE-CNRS UMR8028, Observatoire de Paris, UPMC, 77 Av. Denfert-Rochereau, 75014 Paris, France}

\date{Received xxx ; accepted xxx}

\abstract{A large proportion of observed planetary systems, which contain several planets in a compact orbital configuration, often harbor at least one close-in object. 
In this case, these systems are most likely tidally evolving. We investigate how the effects of planet-on-planet interactions influence the tidal evolution of planets.

To achieve this, we introduced a new open-source addition to the \mercp N-body code, \mTp, which takes  tides, general relativity (GR), and the effect of rotation-induced flattening into account to simulate the dynamical and tidal evolution of multiplanet systems. This code uses a standard equilibrium tidal model, the constant time lag model. 
Additionally, the evolution of the radius of several host bodies has been implemented (e.g., brown dwarfs, M dwarfs of mass $0.1~\Msun$, Sun-like stars, and Jupiter). 
We validate the new code by comparing its output for one-planet systems to the secular equations results. We find that this code  respects the conservation of total angular momentum. 

We then applied this new tool to the planetary system Kepler-62. 
As a result, we find that, in some cases, tides influence the stability of the system. We also show that, while the four inner planets of the systems are likely to have slow rotation rates and small obliquities, the fifth planet could have a fast rotation rate and a high obliquity. 
This means that the two habitable zone planets of this system, Kepler-62e and Kepler-62f, are likely to have very different climate features and of course, this influences their potential for hosting surface liquid water.}
\keywords{Planets and satellites: dynamical evolution and stability -- Planet-star interactions -- Planets and satellites: terrestrial planets -- System: Kepler-62}

\maketitle

\section{Introduction}

More than 1400 exoplanets have now been detected and about 20~\% of them are part of multiplanet systems (\url{http://exoplanets.org/}). 
Many of these systems are compact and host close-in planets where tides have an influence. 
In particular, tides can have an effect on the eccentricities of planets, and also on their rotation periods and their obliquities, which are important parameters for any climate study. 
Moreover, tides can influence the stability of multiplanet systems, as a result of their effect on both the planet's eccentricities and precession rates.

We present a new code, \mTp\footnote{The link to this code and the manual can be found here: \url{http://www.emelinebolmont.com/}.}, which is based on the N-body code \mercp \citep{Chambers1999}. 
This allows us to calculate the evolution of semi-major axis, eccentricity, inclination, rotation period, and obliquity of  planets, as well as the rotation period evolution of the host body. 
This code is flexible, in that it allows us to compute the tidal evolution of systems orbiting any non-evolving object (provided we know its mass, radius, dissipation factor, and rotation period), as well as evolving brown dwarfs (BDs), an evolving M dwarf of $0.1~\Msun$, an evolving Sun-like star, and an evolving Jupiter. 

The dynamics of multiplanet systems with tidal dissipation have been the subject of study (evolution of the orbit in \citealt{WuGoldreich2002,Mardling2007,Batygin2009,Mardling2010,LaskarBoueCorreia2012} and also of the spin in \citealt{WuMurray2003,FabryckyTremaine2007,Naoz2011,Correia2012}), but most of these studies use averaging,  do not study the influence of an evolving host body radius, and  often consider only coplanar systems. 
In this paper, we introduce a tool that allows for more complete studies. 
Indeed, the tidal equations used in this code are not averaged equations, which makes  it  possible to study phenomena such as resonance crossing or capture.
Contrary to other codes using semi-averaged \citep{MardlingLin2002} or non-averaged equations (\citealt{ToumaWisdom1998,MardlingLin2002,Laskar2004}; \citealt{Fienga2008,BeaugeNesvorny2012,CorreiaRobutel2013,MakarovBerghea2013}; and \citealt{Plavchan2015}), our code is freely accessible and open source.

After describing the tidal model used here, we seek to validate the code by comparing one planet's evolutions around BDs with evolutions computed with a secular code (as in \citealt{Bolmont2011,Bolmont2012}). 
We use systems around BDs to test systems where tides are very strong and  lead to important orbital changes. 
We then offer a glimpse of possible research that could use our code and illustrate this with the example of the dynamical evolution of the Kepler-62 system \citep{Borucki2013}.


\section{Model description}
\label{model1}

The major difference between \mercp \ and \mTp is the addition of tidal forces and torques. 
However, we also add the effect of general relativity and rotation-induced deformation.
In the following sections we explain  how these effects were incorporated in the code. We also give the planets and star/BD/Jupiter parameters which are implemented in the code.  

\subsection{Tidal model}
\label{tide_equ}

To compute the tidal interactions, we used the tidal force as expressed in \citet{Mignard1979}, \citet{Hut1981}, \citet{EKH1998}, and \citet{Leconte2010} for the constant time lag model.
This model is based on the assumption that the bodies under review are made of a weakly viscous fluid \citep{Alexander1973}.

We added this force in the N-body code \mercp \citep{Chambers1999}. 
We also consider the tidal forces between the star and the planets but ignore the tidal interaction between planets. 
In addition, we consider here a population of N planets orbiting a star. 

\begin{figure}[htbp!]
        \begin{center}
        \includegraphics[width=9cm]{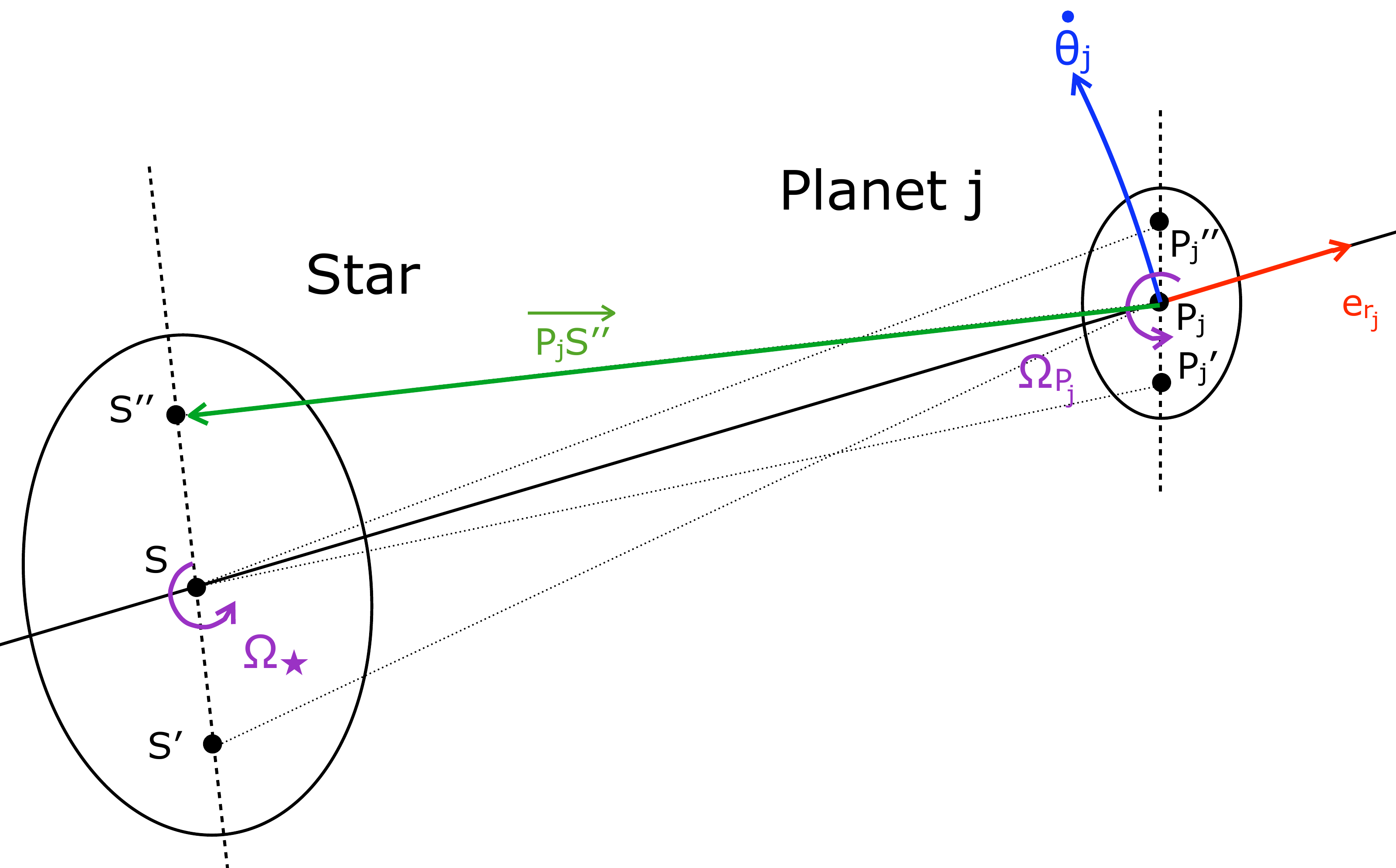}
        \caption{Two-dimensional diagram representing the two deformed bodies. The star is divided into three masses: a central mass of $\Ms-2\mus$ at S, and two bulges of mass $\mus$ at S' and S''. The planet j is divided into three masses: a central mass of $\Mpj-2\mupj$ at $\mathrm{P_j}$, and two bulges of mass $\mupj$ at $\mathrm{P_j}'$ and $\mathrm{P_j}''$. $\mathbf{\Os}$ is the star rotation vector (its norm is $\Os$, the star rotation frequency), $\mathbf{\Opj}$ is planet j rotation vector (its norm is $\Opj$, the planet rotation frequency), and $\dot{\boldsymbol{\theta}_{\rm j}}$ is a vector collinear with the orbital angular momentum of planet j (its norm is equal to the derivative of the true anomaly). $\mathbf{e_{\rj}}$ is the radial vector.}
        \label{schema}
        \end{center}
        \end{figure}

As in \citet{Hut1981}, in order to obtain the expression of the force, we stop the development at the quadrupole order. At this order, we can use the point mass description of the tidal bulges. 
Star and planets are deformed.
Owing to the presence of planet j, the star of mass $\Ms$ is deformed and can be decomposed in a central mass $\Ms-2\mus$, and 2 bulges of mass $\mus$. 
As in \citet{Hut1981}, each bulge is located at a radius $\Rs$ from the center of the star and they are diametrically opposed.
Figure \ref{schema} shows the geometrical context of the problem. The central mass of the star is labeled  S, and the bulges  S' and S''.
The mass of a bulge depends on the time lag and is given by
\begin{equation}\label{mubd}
\mus = \frac{1}{2}\ks\Mpj \Rs^3\left(\rj(t-\taus)\right)^{-3},
\end{equation}
where $\rj$ is the distance between the star and planet j at time $t-\taus$, $\Rs$ is the radius of the star, $\ks$ its potential Love number of degree 2, and $\taus$ its constant time lag.

Because of the presence of the star, the planet j is deformed and can be decomposed in a central mass $\Mpj-2\mupj$, and two bulges of mass $\mupj$. 
The central mass of the planet j is labeled by P$_{\rm j}$, and the bulges by P$_{\rm j}$' and P$_{\rm j}$''.
The bulge's mass is given by 
\begin{equation}\label{mupj}
\mupj = \frac{1}{2}\kpj\Ms \Rpj^3\left(\rj(t-\taupj)\right)^{-3},
\end{equation}
where $\Rpj$ is the radius of planet j, $\kpj$ its potential Love number of degree 2, and $\taupj$ its time lag. To the lowest order in $\taupj$, Equation \ref{mupj} becomes: 
\begin{equation}\label{mu2}
\mupj = \frac{1}{2}\kpj\Ms \left(\frac{\Rpj}{\rj}\right)^{3}\left(1+3\frac{\dot{\rj}}{\rj}\taupj\right).
\end{equation}

Up to the third order in $\Rpj/\rj$ and $\Rs/\rj,$ the forces exerted by the primary on the secondary are the following gravitational forces: $\mathbf{f}_{{\rm S}\rightarrow {\rm P_j}}$, $\mathbf{f}_{{\rm S}\rightarrow {\rm P_j'}}$, $\mathbf{f}_{{\rm S}\rightarrow {\rm P_j''}}$, $\mathbf{f}_{{\rm S'} \rightarrow {\rm P_j}}$, and $\mathbf{f}_{{\rm S''}\rightarrow {\rm P_j}}$, where the latter expression is given by
\begin{equation}\label{force}
\mathbf{f}_{{\rm S''}\rightarrow {\rm P_j}} = \frac{\G\mus(\Mpj-2\mupj)}{\|{\rm P_jS''}\|^3} \mathbf{P_j S''},
\end{equation}
where $\mathbf{P_j S''}$ is the vector $\overrightarrow{P_j S''}$, defined in Figure \ref{schema}.

Let us define $F_{{\rm tr}}$ (for tides radial), $P_{{\rm to},\star}$, and $P_{{\rm to},\pj}$ (for tides ortho-radial) as\begin{equation}\label{F_things}
\begin{split}
F_{{\rm tr}} & =  \frac{-3\G}{\rj^7}\left(\Mpj^2\ks\Rs^5+\Ms^2\kpj\Rpj^5\right) \\ 
& \quad -9\G\frac{\dot{\rj}}{\rj^8}\left(\Mpj^2\Rs^{5}\ks \taus+\Ms^2\Rpj^{5}\kpj \taupj \right),\\
P_{{\rm to},\pj} & = 3\G\frac{\Ms^2\Rpj^{5}}{\rj^7}\kpj\taupj,\\
P_{{\rm to},\star} & = 3\G\frac{\Mpj^2\Rs^{5}}{\rj^7}\ks\taus.
\end{split}
\end{equation}
Here, $F_{{\rm tr}}$ has the dimension of a force (M.L.T$^{-2}$), while $P_{{\rm to},\pj}$ and $P_{{\rm to},\star}$ have a dimension of a momentum (M.L.T$^{-1}$).

Consequently, the total resulting force as a result of the tides acting on planet j is 
\begin{equation}\label{acc}
\begin{split}
\mathbf{F^{T}_{\pj}} &= \left[F_{{\rm tr}} + \left(P_{{\rm to},\star}+P_{{\rm to},\pj}\right)\frac{\mathbf{\vj}.\mathbf{e_{\rj}}}{{\rj}}\right]\mathbf{e_{\rj}} \\
& \quad +  P_{{\rm to},\pj}\left(\mathbf{\Opj}- \dot{\boldsymbol{\theta}_{\rm j}}\right)\times \mathbf{e_{\rj}} \\
& \quad +  P_{{\rm to},\star}\left(\mathbf{\Os} - \dot{\boldsymbol{\theta}_{\rm j}} \right) \times \mathbf{e_{\rj}},
\end{split}
\end{equation}
where $\mathbf{\Os}$ is the star rotation vector, $\mathbf{\Opj}$ is planet j rotation vector, and $\mathbf{\vj}=\mathbf{\dot{r}_{\rm j}}$ is the velocity of planet j. 
The unit vector, $\mathbf{e_{\rj}}$, is defined as $\mathbf{SP_{\rm j}}/\mathrm{SP_{\rm j}}$, while $\dot{\boldsymbol{\theta}_{\rm j}}$ is a vector collinear with the orbital angular momentum of planet j (defined hereafter as $\mathbf{L_{horb}}$), the norm of which is equal to the derivative of the true anomaly.
The term $\dot{\boldsymbol{\theta}_{\rm j}}\times\mathbf{e_{\rj}}$ can be re-written as follows:
\begin{equation}\label{vtheta}
\dot{\boldsymbol{\theta}_{\rm j}}\times\mathbf{e_{\rj}} =\frac{1}{\rj^2}\left(\mathbf{\rj}\times\mathbf{\vj}\right)\times\mathbf{e_{\rj}}.
\end{equation}

What modifies the spin of the star is the following torque contribution: $-\mathbf{\rj}\times\left(\mathbf{f}_{P_{\rm j}\rightarrow S'}+\mathbf{f}_{P_{\rm j}\rightarrow S''}\right)$. 
To calculate the torque on the star, $\mathbf{N^T_{p_{\rm j}\rightarrow \star}}$, we consider the planet as a point mass, meaning that we ignore the gravitational interaction between the bulges of the planet and the bulges of the star. 
Following a similar hypothesis, we find that the torque contribution that modifies the spin of planet j, $\mathbf{N^T_{\star\rightarrow p_{\rm j}}}$, is $\mathbf{\rj}\times\left(\mathbf{f}_{S\rightarrow P_{\rm j}'}+\mathbf{f}_{S\rightarrow P_{\rm j}''}\right)$, so the torque exerted by planet j on the star is given by
\begin{equation}
\mathbf{N^T_{p_{\rm j}\rightarrow \star}} = \mathbf{N^T_{\star}} = P_{{\rm to},\star}\left(\rj~\mathbf{\Os}- \left(\mathbf{\rj}.\mathbf{\Os}\right)\mathbf{e_{\rj}} - \mathbf{e_{\rj}}\times \mathbf{\vj}\right),
\end{equation}
and the torque exerted by the star on the planet j is
\begin{equation}\label{torquep}
\mathbf{N^T_{\star\rightarrow p_{\rm j}}} = \mathbf{N^T_{p_{\rm j}}} = P_{{\rm to},\pj}\left(\rj~\mathbf{\Opj}- \left(\mathbf{\rj}.\mathbf{\Opj}\right)\mathbf{e_{\rj}} - \mathbf{e_{\rj}}\times \mathbf{\vj}\right).
\end{equation}

With this description of the phenomenon, we consider that each planet creates an independent tidal bulge on the star and that the bulge created by planet j does not affect planet i$\neq$j.


\subsection{General relativity}
\label{GR_force}

We added the force due to general relativity as given in \citet{Kidder1995,MardlingLin2002} to \mercp. 
This force corresponds to the orbital acceleration due to the Post-Newtonian potential and its expression is  
 \begin{equation}\label{F_GR}
\begin{split}
\mathbf{F^{GR}_{\pj}}&= \Mpj \left(F_{{\rm GRr}} \mathbf{e_{\rj}}+F_{{\rm GRo}} \mathbf{\vj}\right).
\end{split}
\end{equation}

Here, $F_{{\rm GRr}}$ and $F_{{\rm GRo}}$ are given by
\begin{equation}\label{F_GR_1}
\begin{split}
F_{{\rm GRr}} &= -\frac{\G(\Ms+\Mpj)}{\rj^2 c^2} \\
& \quad \times \left((1+3\eta)\vj^2-2(2+\eta)\frac{\G(\Ms+\Mpj)}{\rj}-\frac{3}{2}\eta\dot{\rj}^2\right) \\
F_{{\rm GRo}} &=  2(2-\eta)\frac{\G(\Ms+\Mpj)}{\rj^2 c^2}\dot{\rj},
\end{split}
\end{equation}
where $\vj$ is the norm of the velocity $\mathbf{\vj}$ of the planet, and $c$ is the speed of light, and where $\eta$ is
\begin{equation}\label{eta}
\eta = \frac{\Ms\Mpj}{(\Ms+\Mpj)^2}.
\end{equation}


\subsection{Rotational deformation}
\label{rot_equ}

The equilibrium figure of a viscous body in rotation is a triaxial ellipsoid symmetric with respect to the rotation axis \citep{MurrayDermott1999}. The rotational deformation is quantified by the parameter $J_2$, defined for
planet j as follows : 
\begin{equation}\label{J2p}
J_{2,{\rm p_j}} =  \kfpj \frac{\Opj^2\Rpj^3}{3\G\Mpj},
\end{equation}
and for the star as
\begin{equation}\label{J2s}
J_{2,\star} =  \kfs \frac{\Os^2\Rs^3}{3\G\Ms}.
\end{equation}
Here, $\kfpj$ is the fluid Love number of planet j and $\kfs$ that of the star.
We define the fluid Love number as the potential Love number for a perfectly fluid planet \citep[see, for example, Figure 2 of][for the Earth's potential Love number and fluid Love number]{CorreiaRodriguez2013}.
Our code allows the user to choose different values for the fluid Love number $\kfp$ and the potential Love number $\kp$.

The total resulting force due to the rotational deformation of star and planet j on planet j is \citep{MurrayDermott1999,Correia2011} 
\begin{equation}\label{acc_rot}
\begin{split}
\mathbf{F^{R}_{\pj}} &= \left(-\frac{3}{\rj^5}\left(C_\star + \Cpj\right)+ \frac{15}{\rj^7}\left(C_\star \frac{\left(\mathbf{\rj}.\mathbf{\Os}\right)^2}{\Os^2} + \Cpj \frac{\left(\mathbf{\rj}.\mathbf{\Opj}\right)^2}{\Opj^2}\right)\right)\mathbf{\rj} \\
& \quad - \frac{6}{\rj^5}\left(C_\star \frac{\mathbf{\rj}.\mathbf{\Os}}{\Os^2}\mathbf{\Os}+\Cpj \frac{\mathbf{\rj}.\mathbf{\Opj}}{\Opj^2}\mathbf{\Opj}\right),
\end{split}
\end{equation}
where $C_\star$ and $\Cpj$ are defined as follows: 
\begin{equation}\label{C_rot}
\begin{split}
C_\star &= \frac{1}{2}\G\Mpj\Ms J_{2,{\rm p_j}}\Rpj^2 \\
\Cpj &= \frac{1}{2}\G\Mpj\Ms J_{2,\star}\Rs^2.
\end{split}
\end{equation}

The torque exerted by planet j on the star is given by
\begin{equation}\label{torquesR}
\mathbf{N^R_{p_{\rm j}\rightarrow \star}} = \mathbf{N^R_{\star}} = - \frac{6}{\rj^5}C_\star \frac{\mathbf{\rj}.\mathbf{\Os}}{\Os^2}\left(\mathbf{\rj}\times\mathbf{\Os}\right),
\end{equation}
and the torque exerted by the star on the planet is 
\begin{equation}\label{torquepR}
\mathbf{N^R_{\star\rightarrow p_{\rm j}}} = \mathbf{N^R_{p_{\rm j}}} = - \frac{6}{\rj^5}\Cpj \frac{\mathbf{\rj}.\mathbf{\Opj}}{\Opj^2}\left(\mathbf{\rj}\times\mathbf{\Opj}\right).
\end{equation}

These torques are responsible for the precession of the orbit normal. 
This precession has an influence on the mean eccentricity of  planets and also on their mean obliquity.

\subsection{Summary of all the effects}

\subsubsection{Corrective acceleration}

To compute the evolution of the planetary orbits, we need to take all the resulting effects into account.
The orbital part of the acceleration is handled by \mercp, so below we provide the expression of the corrective acceleration of planet j in the astrocentric coordinates:
\begin{equation}\label{acceleration_tot}
\begin{split}
\mathbf{a}_{{\rm p}_{\rm j}} &= \frac{\Mpj+\Ms}{\Mpj\Ms}\left(\mathbf{F^{T}_{\pj}}+\mathbf{F^{GR}_{\pj}}+\mathbf{F^{R}_{\pj}}\right) + \frac{1}{\Ms} \sum\limits_{i\ne j}^N \left(\mathbf{F^{T}_{\pli}}+\mathbf{F^{GR}_{\pli}}+\mathbf{F^{R}_{\pli}}\right)\\
& = \frac{1}{\Mpj}\left(\mathbf{F^{T}_{\pj}}+\mathbf{F^{GR}_{\pj}}+\mathbf{F^{R}_{\pj}}\right) + \frac{1}{\Ms} \sum\limits_{i=1}^N \left(\mathbf{F^{T}_{\pli}}+\mathbf{F^{GR}_{\pli}}+\mathbf{F^{R}_{\pli}}\right),
\end{split}
\end{equation}
where $\mathbf{F^{T}_{\pj}}$, $\mathbf{F^{GR}_{\pj}}$ and $\mathbf{F^{R}_{\pj}}$ are defined in the previous sections.

\subsubsection{Spin equations}

First, let us consider a system with one planet.  
We hypothesize that we can decouple the torque equation given by the conservation of total angular momentum $\mathbf{L }$. 
This equation is the following:
\begin{equation}\label{mom}
\begin{split}
& \frac{\d}{\d t} \mathbf{L}  = \mathbf{0}, \\
& \frac{\d}{\d t}\left(I_{\star}\mathbf{\Os} + \Ip \mathbf{\Op}+ \mathbf{L_{orb}}\right) = \mathbf{0},
\end{split}
\end{equation}
where $I_{\star}$ is the principal moment of inertia of the star, $\Ip$ that of the planet and $\mathbf{ L_{orb}}$ is the orbital angular momentum: 
\begin{equation}\label{horb1}
\mathbf{L_{orb}} =\mathbf{r^\ast}\times\Mp \mathbf{v^\ast} + \mathbf{r_\star^\ast}\times\Ms \mathbf{v_{\star}^\ast},
\end{equation}
where $\mathbf{r^\ast}$ and $\mathbf{v^\ast}$ are the position and velocity of the planet in the reference frame of the center of mass of the system.
The position and velocity of the star in the reference frame of the center of mass of the system are respectively $\mathbf{r_{\star}^\ast}$ and $\mathbf{v_{\star}^\ast}$.

In astrocentric coordinates, the position and velocity of the planet are $\mathbf{r}$ and $\mathbf{v}$ and Equation \ref{horb1} becomes:  
\begin{equation}\label{horb2}
\mathbf{L_{orb}} =  \frac{\Ms\Mp}{\Ms+\Mp}\mathbf{r}\times \mathbf{v},
\end{equation}
so 
\begin{equation}\label{dhorb}
\begin{split}
\frac{\d}{\d t}\bigg(\mathbf{L_{orb}}\bigg) &= \frac{\Ms\Mp}{\Ms+\Mp}\frac{\d}{\d t}\bigg(\mathbf{r}\times \mathbf{v}\bigg)\\
&=  \frac{\Ms}{\Ms+\Mp}\left(\mathbf{N_{\star\rightarrow p}}+\mathbf{N_{p\rightarrow \star}}\right),
\end{split}
\end{equation}
where $N_{\star\rightarrow p}$ is the total torque exerted on the planet, and $N_{p\rightarrow \star}$  the total torque exerted on the star.

Assuming that the spins of the star and the planet evolve solely as a result of tidal and rotational flattening torques, this means that we can decouple Equation \ref{mom} to obtain the following spin equations:
\begin{equation}\label{spinequations1}
\begin{cases}
 \frac{\d}{\d t}\left(I_{\star}\mathbf{\Os} \right) & = -\frac{\Ms}{\Ms+\Mp} \left(\mathbf{N^T_{\star}}+ \mathbf{N^R_{\star}}\right)\\
 \frac{\d}{\d t}\bigg(\Ip \mathbf{\Op}\bigg) & = -\frac{\Ms}{\Ms+\Mp} \left(\mathbf{N^T_{p}}+ \mathbf{N^R_{p}}\right),
\end{cases}
\end{equation} 

If there is more than one planet in the system, the orbital angular momentum involves planet j-- planet i$\neq$j cross terms (e.g., in $\mathbf{\rj}\times\mathbf{\vi}$, $\mathbf{\ri}\times\mathbf{\vj}$) that we ignore for the spin calculation.
The equations governing the evolution of the spin of the star and the spin of planet j are, therefore, 
\begin{equation}
\begin{cases}
 \frac{\d}{\d t}\left(I_{\star}\mathbf{\Os} \right) & = -\sum\limits_{j=1}^N \frac{\Ms}{\Ms+\Mpj} \left(\mathbf{N^T_{\star}}+ \mathbf{N^R_{\star}}\right)\\ \label{spin_tot1}
 \frac{\d}{\d t}\bigg(\Ipj \mathbf{\Opj}\bigg) & = -\frac{\Ms}{\Ms+\Mpj} \left(\mathbf{N^T_{p_{\rm j}}}+ \mathbf{N^R_{p_{\rm j}}}\right),
\end{cases}
\end{equation} 
where $\mathbf{N^T_{\star}}$, $\mathbf{N^R_{\star}}$, $\mathbf{N^T_{p_{\rm j}}}$, and $\mathbf{N^R_{p_{\rm j}}}$ are defined in the previous sections.
The spin of planet j is the norm of the vector $\mathbf{\Opj}$ and the spin of the star is the norm of the vector $\mathbf{\Os}$.

Our \mTp code provides the x, y, and z components of the spin of the planets, of their orbital angular momentum, and of the spin of the star. 
Then the obliquity $\oblpj$ and the inclination $i_{\rm j}$ of planet j are obtained by calculating
\begin{equation}\label{obl_spin_def}
\begin{split}
\cos\oblpj &= \frac{\mathbf{L_{{orb}_{\rm j}}} \cdot \mathbf{\Opj}}{\left\Vert \mathbf{L_{{orb}_{\rm j}}} \right\Vert \times \left\Vert \mathbf{\Opj} \right\Vert} \\
\cos i_{\rm j} &= \frac{\mathbf{L_{{orb}_{\rm j}}} \cdot \mathbf{\Os}}{\left\Vert \mathbf{L_{{orb}_{\rm j}}} \right\Vert \times \left\Vert \mathbf{\Os} \right\Vert},
\end{split}
\end{equation}
 where $\mathbf{L_{{orb}_{\rm j}}}$ is a vector normal to the orbit of planet j.

\subsection{Integration of the spin}

In this study, we use the \mercp code's hybrid routine that relies on a Hamiltonian description of the problem. 
The Hamiltonian is divided into three parts that are integrated consecutively \citep{Chambers1999}. 
\mercp allows the user to add other forces on the planets via a routine. 
In the Hamiltonian description, these extra forces are treated as a perturbation to the Keplerian potential.
The user should keep in mind that this violates the symplectic properties of the integrator.

\begin{figure}[htbp!]
        \begin{center}
        \includegraphics[width=\linewidth]{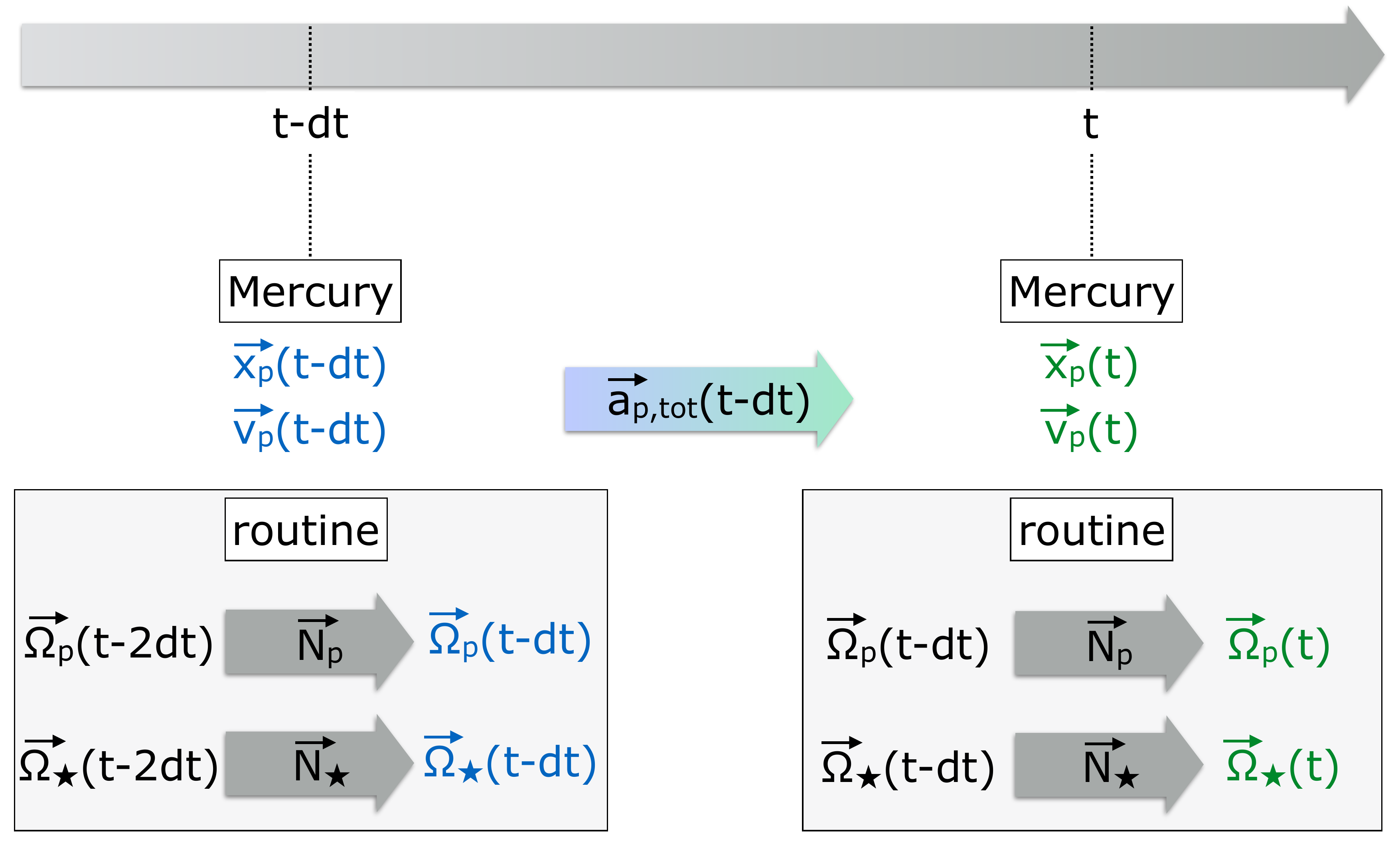}
        \caption{Integration scheme of \mTp.}
        \label{schema_integration}
        \end{center}
        \end{figure}

In this user routine, we add the tidal forces, rotation-induced flattening forces and GR forces. 
On the one hand, we computed the resulting acceleration on the planets, and this acceleration is used by the bulk of \mercp to compute the evolution of the system. 
On the other hand, the spin of  planets and the star is integrated with a 5th order Runge-Kutta within the routine. 
The Runge-Kutta integration is performed twice in a \mercp time step. 
To compute the spins at a time $t$, the positions and velocities of the planets are interpolated between $t-$d$t$ and $t$ at the time intervals required for the Runge-Kutta routine.
The integration scheme is illustrated in Figure \ref{schema_integration}.

If the host body evolves, the moment of inertia of the star, given by: $I_{\star} = \Ms (rg_{\star}\Rs)^2$, where $rg_{\star}$ is the radius of gyration \citep{Hut1981}, varies with time.

The equation of each component of the spin $\mathbf{\Os}$ is given in the following equation, written here for the z component $\Osz$:  
\begin{equation}\label{spin_R_varies}
\begin{split}
I_{\star}(t)\Osz(t) & = I_{\star}(t-\d t)\Osz(t-\d t) \\
& \quad - \int_{t-\d t}^{t} \sum\limits_{j=1}^N \frac{\Ms}{\Ms+\Mpj}\left(N^T_{\star, z}+ N^R_{\star, z}\right)\d t.
\end{split}
\end{equation}

The error introduced by the spin integration depends on the third power of the time step \citep[][]{Chambers1999}. 
The part of the integration that causes more errors is the integration of the rotation-induced flattening effect. 
The integration of the tidal torque does not require such a precise integration owing to long timescales of evolution with respect to the \mercp time step (which is usually taken as slightly smaller than one tenth of the inner planet's orbital period). 
However, the rotation-induced flattening causes changes in the spin of the planets on a much shorter timescale. 

For very close-in planets, the integration of the rotation-induced flattening effect can lead to a purely numerical decrease in the rotation period. 
It is therefore important to evaluate the error made during the integration of the rotation-induced flattening torque.
We suggest using a time step shorter than the orbital period of the inner planet divided by 20. 
The time step should therefore be chosen according to the precision required in the rotation period of the inner planet (see section \ref{rot_flat} for details).


\subsection{Input parameters}

\begin{table*}[htbp]
\begin{center}
\caption{Planetary parameters implemented in \mTp.}
\vspace{0.1cm}
\begin{tabular}{c|c|c|c|c|c|c}
Type of planet  & Mass & Radius & Love  & Moment of inertia & \multicolumn{2}{|c}{Time lag} \\
 & & & number & $I/(MR^2)$ & Value (s) & Notation \\
\hline
Earth-mass planet & 1~$\Mearth$ & 1~$\Rearth$ & 0.305 & 0.3308 & $698$ & $\tearth$ \\
Jupiter & 1~$\Mjup$ & 1~$\Rjup$ & 0.380 & 0.254 & $1.842\times10^{-3}$ & $\tauHJ$ \\
\end{tabular} 
\label{tab:param:planets} 
\end{center}
\end{table*}

\begin{table*}[htbp]
\begin{center}
\caption{Host body parameters implemented in \mTp.}
\vspace{0.1cm}
\begin{tabular}{c|c|c|c|c|c|c}
Type of   & Mass & Radius & Love  & Moment of inertia & \multicolumn{2}{|c}{Dissipation factor}    \\
 host body & & & number & $I/(MR^2)$ & Value (g$^{-1}$cm$^{-2}$s$^{-1}$) & Notation  \\
\hline
Jupiter & 1~$\Mjup$ & \textbf{evolving} & \textbf{evolving} & \textbf{evolving} & $7.024 \times 10^{-59}$ & $\sigma_{\rm J}$  \\
BD & 0.01--0.08~$\Msun$ & \textbf{evolving} & 0.379--0.307 & \textbf{evolving} & $2.006 \times 10^{-60}$ & $\sbd$ \\
dM & 0.1~$\Msun$ & \textbf{evolving} & 0.307 & 0.2 & $2.006 \times 10^{-60}$ & $\sigma_{{\rm d}M}$  \\
Sun & 1~$\Msun$ & \textbf{evolving} & 0.03 & 0.059 & $4.992 \times 10^{-66}$ & $\sigma_\odot$  \\
\end{tabular} 
\label{tab:param:stars} 
\end{center}
\end{table*}

The code requires the necessary planetary parameters to work and these are used in all the equations in sections \ref{tide_equ}, \ref{GR_force}, and \ref{rot_equ}. 
An N-body integrator requires parameters such as: the masses of the planets $\Mpj$, the mass of the host body $\Ms$, the semi-major axis (SMA) of the planets, their eccentricities (ecc), their inclination (inc), and their orbital angles (argument of pericenter, longitude of the ascending node, and mean anomaly). 
In the following tests, the orbital angles are set to 0$\deg$.

To calculate the evolution that results from rotational flattening, the fluid Love number of the star $\kfs$ and of the planets $\kpj$ is required.
To calculate the tidal evolution, the following are also required:  the radius of the star $\Rs$ and the radius of the planets $\Rpj$; the potential Love number of degree two of the star $\ks$ and of the planets $\kpj$; the time lag of the star $\taus$ and of the planets $\taupj$.

All  these parameters are, of course, changeable in our code, but we implemented some useful values and relations for ease of use. These implemented values are given in Tables \ref{tab:param:planets} and \ref{tab:param:stars}. 

\subsubsection{Planet model}

\begin{table*}[htbp]
\begin{center}
\caption{Values of the reduced dissipation factor $\overline{\ss}$ for the host bodies implemented in \mTp.}
\vspace{0.1cm}
\begin{tabular}{c|c|c|c|c|c}
Type of   & Mass & \multicolumn{2}{|c}{Radius}  & \multicolumn{2}{|c}{$\overline{\ss}$}   \\
 host body & & at $t= 1$~Myr & at $t= 1$~Gyr & at $t= 1$~Myr & at $t= 1$~Gyr \\
\hline
Jupiter & 1~$\Mjup$ & $0.15~\Rsun$ & $0.10~\Rsun$ & $4.3\times10^{-5}$ & $1.1\times10^{-5}$  \\
BD & 0.01~$\Msun$ & $0.29~\Rsun$ & $0.10~\Rsun$ & $4.0\times10^{-5}$ & $9.8\times10^{-7}$ \\
        & 0.08~$\Msun$ & $0.85~\Rsun$ &  $0.10~\Rsun$ &$4.9\times10^{-3}$ &$2.7\times10^{-6}$\\
dM & 0.1~$\Msun$ & $0.98~\Rsun$ & $0.12~\Rsun$ &$9.0\times10^{-3}$&$5.8\times10^{-6}$\\
Sun & 1~$\Msun$ & $2.3~\Rsun$& $0.91~\Rsun$ &$1.4\times10^{-6}$&$5.5\times10^{-8}$ \\
\end{tabular} 
\label{tab:param:stars2} 
\end{center}
\end{table*}

\begin{table*}[htbp]
\begin{center}
\caption{Test simulations for tides: one BD and one planet.}
\vspace{0.1cm}
\begin{tabular}{c|c|c|c||c|c|c|c|c|c|c|c|c|c|c|c|}
\multicolumn{1}{c}{} &\multicolumn{3}{c||}{Effects}& \multicolumn{12}{c}{Parameters and initial conditions} \\ 
\hhline{~---------------}
&\multicolumn{2}{c|}{Tides} & BD & $\Ms$ & $\Rs$ & $P_{\star,0}$ & $\Mp$ & SMA & ecc & inc & $P_p$ & $\epsilon_p$ & $\taup$ & $\ss$ & d$t$ \\
\hhline{~--~~~~~~~~~~~~~}
&  BD & Pl. & evol. & ($\Msun$) & ($\Rsun$) & (day) & ($\Mearth$) & (AU) & & (deg) & (hr) & (deg) && & (day)\\
\hline
\textbf{1}  & \xmark & \cmark & \xmark & 0.08 & -- & -- & 1 & 0.014 & 0.1 & 0 & 24 & 11.5 & $\tearth$ & -- & 0.08\\
\textbf{1'}  & \xmark & \cmark & \xmark & 0.08 & -- & -- & 318 & 0.014 & 0.01 & 0 & 24 & 11.5 & 100~$\tauHJ$ & -- & 0.08\\
\hline
\textbf{2}  & \cmark & \xmark & \xmark & 0.08 & 0.85 & 2.9 & 1 & 0.018 & 0 & 11.5 & -- & -- & -- & $1000~\sbd$ & 0.08\\
\hline
\textbf{3}  & \cmark & \cmark & \xmark & 0.08 & 0.85 & 2.9 & 1 & 0.018 & 0.1 & 5 & 24 & 11.5 & $\tearth$ & $\sbd$ & 0.08\\
\textbf{3'} & \cmark & \cmark & \xmark & 0.08 & 0.85 & 2.9 & 318 & 0.009 & 0.1 & 5 & 240 & 40 & 100~$\tauHJ$ & 0.01~$\sbd$ & 0.005\\
\hline
\textbf{4}  & \cmark & \cmark & \cmark & 0.08 & \textbf{evol} & 2.9 & 1 & 0.018 & 0.1 & 0 & 24 & 11.5 & $\tearth$ & $\sbd$ & 0.05\\
\hline
\end{tabular} 
\label{tab:sim:tides} 
\end{center}
\end{table*}

For Earth-like planets and super-Earths, we assume that the product of the potential Love number of degree two with the time lag of the planet is equal to that of  Earth: $k_{2,{\rm p}}\Delta \tau_{\rm p} = k_{2,\oplus}\Delta \tau_{\oplus}$. 
We use here the value of $k_{2,\oplus}\Delta \tau_{\oplus} = 213$~s given by \citet{DeSurgyLaskar1997}. 
We assume here that the fluid Love number and the potential Love number of degree two are equal. 

Given the mass of the planet, we offer the user two possibilities to choose the radius: either it gives a value itself or it assumes a composition and the code calculates the radius following \citet[][]{Fortney2007}. 
For example, a super-Earth of 10~$\Mearth$ would have a radius of 1.8~$\Rearth$.

For a Jupiter-like planet, we computed the time lag $\tauHJ$ from the value of the dissipation parameter $\sigma_{\rm k}$ for hot Jupiters of \citet{Hansen2010}. 
The notation $\sigma_{\rm k}$ was introduced by \citet{EKH1998} and is linked to the quantity $k_{2,{\rm k}}\tau_{\rm k}$ by 
\begin{equation}\label{hansen}
k_{2,{\rm k}}\tau_{\rm k} = \frac{3}{2}\frac{R_{\rm k}^5\sigma_{\rm k}}{\G},
\end{equation}
where k represents either the star or a planet.


\subsubsection{Host body evolution and dissipation}\label{dissipation_stuff}

It is possible to use stellar evolution tracks in \mTp to compute the evolution of planets around an evolving object. 
We implemented this for an evolving Jupiter \citep{LeconteChabrier2013}, and for evolving BDs of masses: 0.01, 0.012, 0.015, 0.02, 0.03, 0.04, 0.05, 0.06, 0.07, 0.072, 0.075, and $0.08~\Msun$ \citep{Leconte2011}, for a M dwarf (dM) of mass $0.1~\Msun$ and a Sun-like star \citep{ChabrierBaraffe1997,Baraffe1998}.

Table \ref{tab:param:stars} shows for each type of host body which ones are the evolving quantities and which ones have implemented values. 
The evolving quantities are tabulated and \mTp interpolates the values during the integration to have the correct radius, Love number, moment of inertia in the acceleration formula and the spin equations.

At this point, we hypothesize that, during their evolution the dissipation factor, $\sigma_{\rm k}$ of the host body remains constant.
We use the value of the dissipation of Jupiter given in \citet{Leconte2010} for the Jupiter host body.
We use the dissipation factor of \citet{Bolmont2011} for BDs and for the M dwarf, and we use the value of stellar dissipation of \citet{Hansen2010}  for the dissipation of the Sun-like star. 

In Table \ref{tab:param:stars2}, we indicate the value of the parameter $\overline{\ss}$ for the different bodies, where $\overline{\ss}$ is defined as
\begin{equation}\label{sigma_bar}
\overline{\ss} = \Ms \Rs^2 P_{\rm ff}\ss,
\end{equation}
and where $P_{\rm ff} = \sqrt{\Rs^3/\G\Ms}$ is the free-fall time at the surface of the star.
As this definition of $\overline{\ss}$ depends on the radius of the star, we compute its value for all the bodies of Table \ref{tab:param:stars} for an age of $1$~Myr and of $1$~Gyr.

\begin{table*}[htbp]
\begin{center}
\caption{Test simulation for rotational flattening: one BD with 2 planets}
\vspace{0.1cm}
\begin{tabular}{c|c|c||c|c|c|c|c|c|c|c|c|c|}
\multicolumn{1}{c}{} &\multicolumn{2}{c||}{Effects}& \multicolumn{10}{c}{Parameters and initial conditions for planets 1 and 2} \\ 
\hhline{~------------}
&\multicolumn{2}{|c||}{Rot. flat.{}} & $\Ms$ & $M_{{\rm p}_{1/2}}$ & SMA$_{1/2}$ & ecc$_{1/2}$ & inc$_{1/2}$ & $P_{{\rm p}_{1/2}}$ & $\epsilon_{{\rm p}_{1/2}}$ & $k_{2,{\rm p}_{1/2}}$ & $\ks$ & d$t$ \\
\hhline{~--~~~~~~~~~~}
& BD & Planets & ($\Msun$) & ($\Mearth$) & (AU) & & (deg) & (hr) & (deg) && & (day)\\
\hline
\textbf{5} & \cmark & \xmark  & 0.08 & 1/1 & 0.018/0.025 & 0.01/0.01 & 0/1 & 24/24 & 11.459/11.459 & -- & 0.307 & 0.05\\
\hline
\textbf{6} & \xmark & \cmark  & 0.08 & 1/1 & 0.018/0.025 & 0.01/0.01 & 0/1 & 24/24 & 11.459/11.459 &  0.305/0 & -- & 0.08\\
\textbf{6'} & \xmark & \cmark  & 0.08 & 1/1 & 0.018/0.025 & 0.01/0.01 & 0/1 & 24/24 & 11.459/11.459 &  0.305/0 & -- & 0.05\\
\textbf{6''} & \xmark & \cmark  & 0.08 & 1/1 & 0.018/0.025 & 0.01/0.01 & 0/1 & 24/24 & 11.459/11.459 &  0.305/0 & -- & 0.01\\
\textbf{6'''} & \xmark & \cmark  & 0.08 & 1/1 & 0.018/0.025 & 0.01/0.01 & 0/1 & 24/24 & 11.459/11.459 &  0.305/0 & -- & 0.001\\
\hline
\textbf{7} & \cmark & \cmark & 0.08 & 1/1 & 0.018/0.025 & 0.01/0.01 & 0/1 & 24/24 & 11.459/23 &  0.305/0.305 & 0.307 & 0.08\\
\hline
\end{tabular} 
\label{tab:sim:rot_flat} 
\end{center}
\end{table*}


\section{Code verification}

To validate the tidal part of the code, we first simulated the tidal evolution of one Earth-mass planet orbiting a $0.08~\Msun$ BD with two different approaches. 
The first approach is to use a secular code that solves the averaged equations of the tidal evolution of one planet \citep[equations in semi-major axis, eccentricity, etc, which is  often used in tidal studies such as][]{Hut1981,Leconte2010,Bolmont2011}. 
The second simulation was performed using the \mTp code we developed. 
We first compared the outcomes for different regimes, switching the planetary tide on or off
 (i.e., the tide raised by the BD in the planet) and the BD tide (i.e., the tide raised by the planet in the BD), and testing the effect of the evolving radius of the BD. The details of all simulations are listed in Table \ref{tab:sim:tides}, where d$t$ is the time step used for the simulation.

To test the rotational flattening part of the code, we compared our results with the numerical code used in \citet{CorreiaRobutel2013}, hereafter denoted by the CR13 code. 
This code was developed independently of the present one and uses the ODEX integrator \citep[e.g.,][]{HairerNorsettWanner1993}, but it has not been made available for public use.
In \citet{CorreiaRobutel2013}, the CR13 code was applied to a specific situation: the spin evolution of trojan bodies, but it is much more general than that. 
The CR13 code has the ability to perform the same kind of simulations as those of the {\it Mercury-T} code.
Therefore, the CR13 code is used for cross-checking some of our results.

We considered a two-planet system orbiting a $0.08~\Msun$ BD and validated the effect of the rotational flattening of the star, and of the planet and compared them with our results for a full simulation (effects of tides and rotational flattening). 
The details of the simulations are listed in Table \ref{tab:sim:rot_flat}, where $k_{2,{\rm p}_{1/2}}$ is the Love number of degree 2 of planets 1 and 2, and $\ks$  is the Love number of degree 2 of the host body. 
        
\subsection{Non-evolving BD: the effect of planetary tide}

This case corresponds to Case \textbf{(1)} of Table \ref{tab:sim:tides}, for which we switched off the effect of the BD tide. 

        \begin{figure*}[htbp!]
        \centering
        \includegraphics[width=15cm]{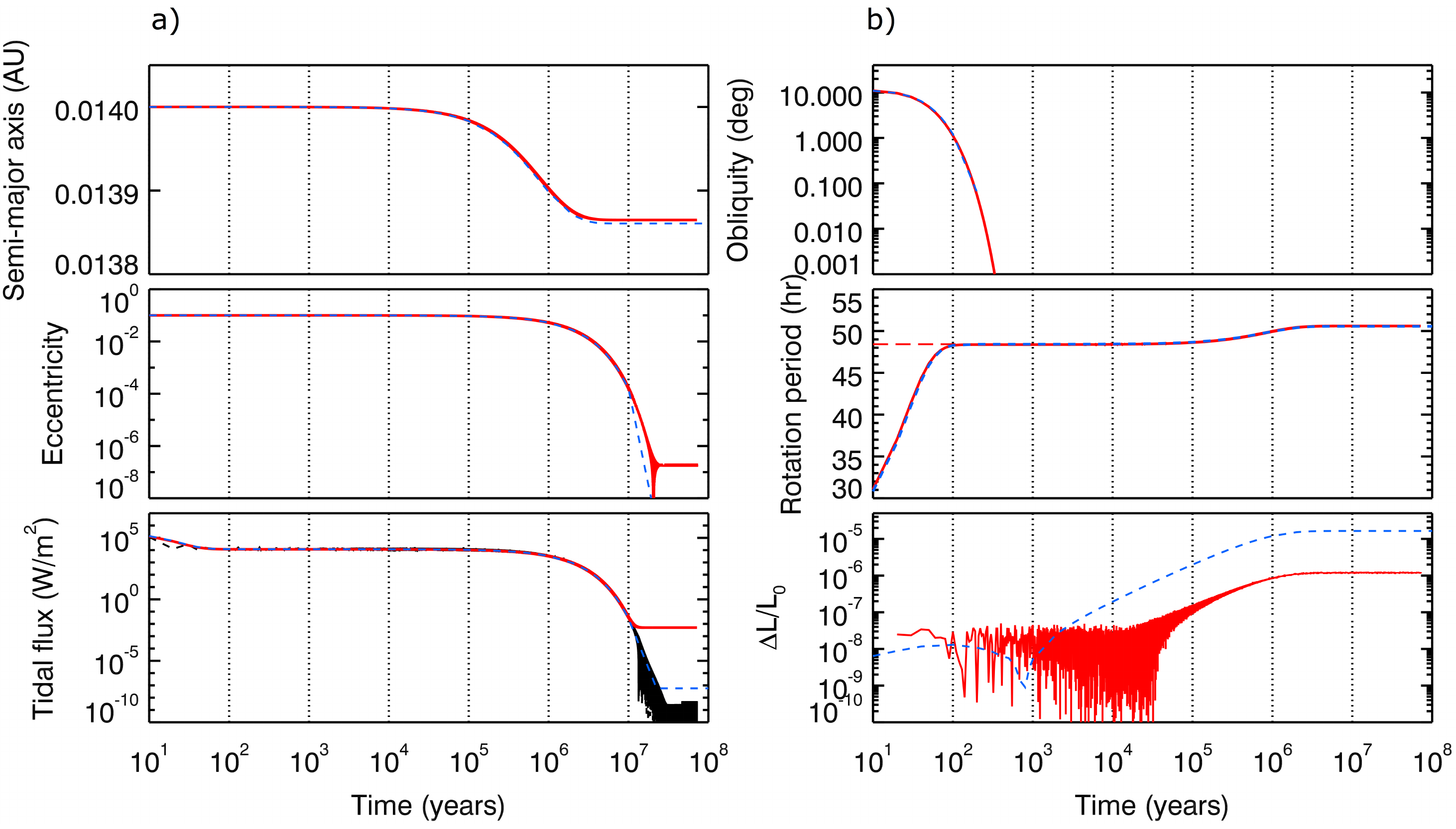}
        \caption{Case \textbf{(1)}: Tidal evolution of a planet of mass $1~\Mearth$ orbiting a  $0.08~\Msun$ BD, calculated with the secular code (blue dashed line) and the \mTp code (solid red line). Graph a) from top to bottom: evolution of the semi-major axis of the planet, evolution of its eccentricity, and evolution of the tidal heat flux. Graph b) from top to bottom: evolution of the obliquity of the planet, evolution of its rotation period (the pseudo-synchronization period represented in long red dashes), and conservation of total angular momentum.}
        \label{12796_aef}
        \end{figure*}
                
Figure \ref{12796_aef} shows the evolution of the semi-major axis, the eccentricity, and the averaged tidal heat flux $\langle \phi_{\textrm{tides}}\rangle$ defined as
\begin{equation}\label{tidalheatflux}
\langle\phi_{\textrm{tides}}\rangle = \langle\dot{E}_{\textrm{tides}}\rangle/4\pi \Rp^2,
\end{equation}
where $\langle\dot{E}_{\textrm{tides}}\rangle$ is the averaged gravitational energy lost by the system by dissipation.
Here,  
\begin{equation}\label{Edot_tides}
\begin{split}
\langle\dot{E}_{\textrm{tides}}\rangle = &2 \frac{1}{\Tp}\frac{\G \Mp \Ms}{4a} \Bigg[Na1(e) - 2Na2(e) \cos \oblp \frac{\Op}{n} \\
&+\left(\frac{1+\cos^2\oblp}{2}\right) \Omega(e) \left(\frac{\Op}{n}\right)^2 \Bigg],
\end{split}
\end{equation}
where $\Tp$ is the dissipation timescale. 
$Na1(e)$, $Na2(e),$ and $\Omega(e)$ are eccentricity-dependent factors defined in \citet{Bolmont2013}. 
The tidal heat flux depends on the eccentricity and on the obliquity of the planet. If the planet has no obliquity and no eccentricity and if its rotation is synchronized, the tidal heat flux is zero. 

We also show the evolution of the instantaneous tidal heat flux, computed from the instantaneous energy loss given by
\begin{equation}\label{Edot_inst}
\dot{E}_{\textrm{tides}}(t) = - \dot{E}_{\textrm{orb}}(t) = -\left(\mathbf{F^{T}_{\pj}}\cdot\mathbf{\vj} + \Ipj \mathbf{\Opj}\cdot \mathbf{\dot{\Omega}_{\pj}}\right),
\end{equation}
where $\mathbf{\dot{\Omega}_{\pj}}$ is the derivative of the spin of planet j, given by Equation \ref{spin_tot1}. 
Contrary to $\langle\dot{E}_{\textrm{tides}}\rangle$ which depends on averaged computed values, such as the semi-major axis and the eccentricity, $\dot{E}_{\textrm{tides}}(t)$ depends on the instantaneous position, velocity, and spin of planet j.

The eccentricity of the planet decreases to values below $10^{-4}$ in $10^7$~yr. 
The decrease of the eccentricity of the planet is accompanied by a decrease of the semi-major axis. 
The evolution of these two quantities shows a good agreement between the secular code and the \mTp code.
The obliquity of the planet decreases from its initial value of $11.5\deg$ to less than $10^{-4}$~degrees in less than $500$~yr. 
During the same time, the rotation period evolves from its initial value of $24$~hr to the pseudo-synchronization period, which in this study is  $\sim 48.5$~hr. 
The evolution of obliquity and rotation period show a good agreement between the secular code and \mTp. 

After $2\times10^7$~yr of evolution, the eccentricity obtained with \mTp is equal to a few $10^{-7}$. 
This residual value of the eccentricity comes from the way the \mercp code calculates the orbital elements from the positions and velocities of the planets. 
Indeed, it assumes a Keplerian potential. 
However, in this situation, where the tidal forces are taken into account, this is not the case.

Nevertheless, we can assume that an eccentricity of $10^{-7}$  be considered as null. 
Furthermore, this code is designed to study multiplanet systems. 
In the examples we give later, the eccentricity due to planet-planet interactions is typically greater than $10^{-7}$. 

This residual eccentricity is responsible for a non-zero-averaged tidal heat flux $\lesssim 10^{-2}$~W/m$^2$. 
However, the instantaneous tidal heat flux reaches values as low as a few $10^{-9}$~W/m$^2$.
This low value illustrates the fact the real eccentricity of the planet must be much lower than that which \mTp is calculating.

        \begin{figure*}[htbp!]
        \centering
        \includegraphics[width=15cm]{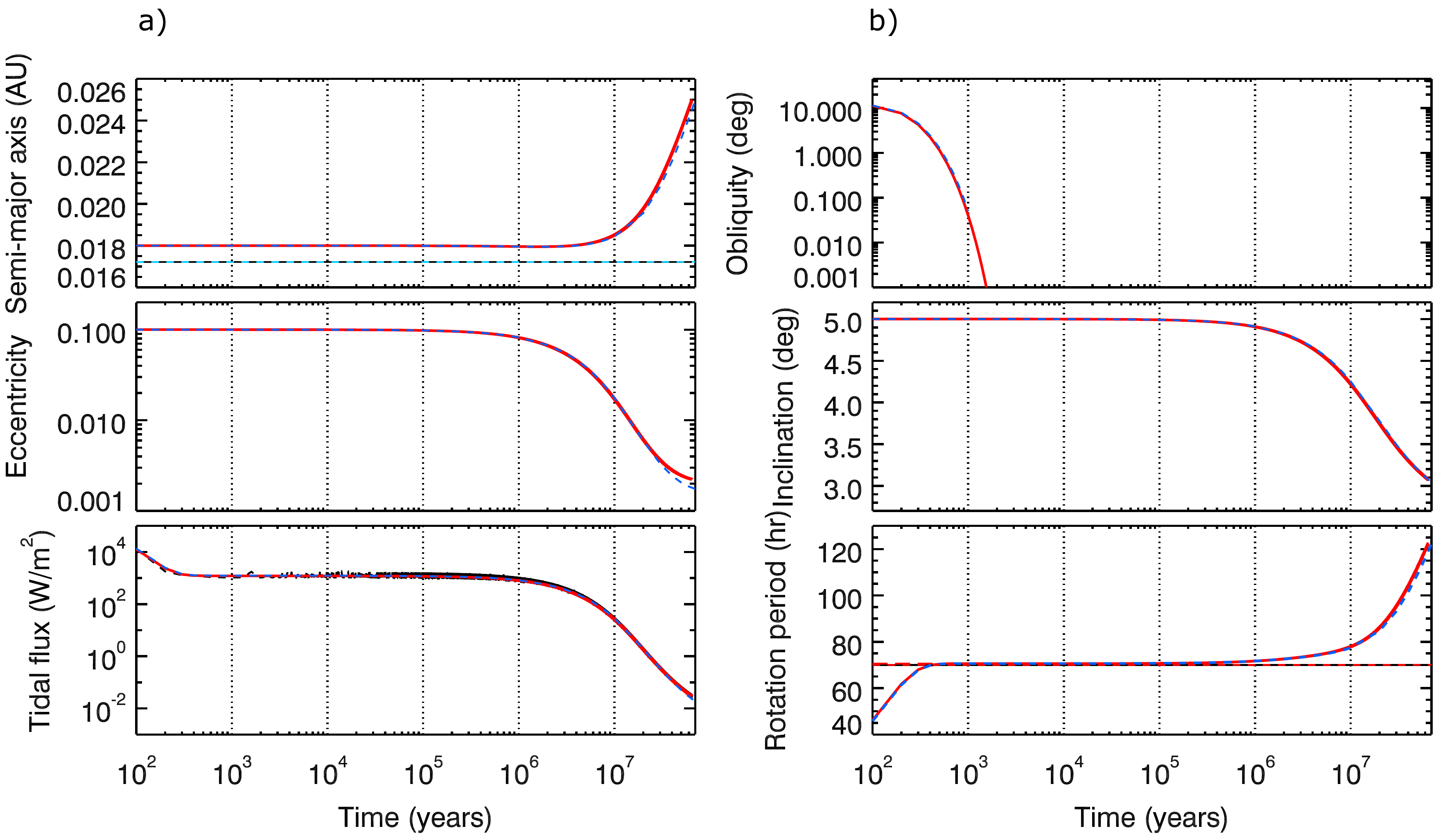}
        \caption{Case \textbf{(3)}: Tidal evolution of a planet of mass $1~\Mearth$ orbiting a  $0.08~\Msun$ BD, calculated with the secular code (blue dashed line) and the \mTp code (solid red line). Graph a) from top to bottom: evolution of semi-major axis (in red) and of the corotation distance (in black), evolution of eccentricity, and evolution of tidal heat flux. Graph b) from top to bottom: evolution of the obliquity of the planet, evolution of its inclination and evolution of its rotation period (in red) and the BD rotation period (in black). The pseudo-synchronization period is represented in long red dashes.}
        \label{12799_aef_ois}
        \end{figure*}

\medskip

We also verify that each component of the total angular momentum is a conserved quantity during the evolution of the system (Equation \ref{mom}). 
We define the quantity $\alpha_i$ where $i$ is x, y, or z, and $\alpha$ as 
\begin{equation}\label{alpha_mom}
\begin{split}
\alpha_i &= \frac{\mathrm{L}_i(t)-\mathrm{L}_i(0)}{\mathrm{L}(0)},\\
\alpha & = \frac{\mathrm{L}(t)-\mathrm{L}(0)}{\mathrm{L}(0)},
\end{split}
\end{equation}
where $\mathrm{L}_i$ is the i component of the total angular momentum vector and $\mathrm{L}$ is the norm of the vector $\mathbf{L}$ of Equation \ref{mom}. 
In this example, we only consider the effect of the planetary tide, which is equivalent to considering the BD as a point mass, so that the total angular momentum, in this case, is  only the sum of the orbital angular momentum and the rotational angular momentum of the planet. 

The bottom right  panel of Figure \ref{12796_aef} shows the conservation of the total angular momentum as a function of time. 
For the \mTp simulation, each component of the total angular momentum $\alpha_i$ is conserved and $\alpha$ reaches $10^{-6}$ after 100~Myr of evolution. 
For the secular code, the total angular momentum is a little less well conserved and reaches a few $10^{-5}$ at the end of the simulation.
In this example, the orbital angular momentum is $10^5$ higher than the angular momentum of the planet, so the question  remains as to whether the spin of the planet has been correctly computed.

To test this, we performed another simulation, Case \textbf{(1')},  with a Jupiter-mass planet to reduce the difference between the orbital angular momentum and the planet's angular momentum. 
In this case, the orbital angular momentum is about $10^3$ higher than the angular momentum of the planet. 
We find that for this simulation the total angular momentum is conserved, with $\alpha$ asymptopically reaching only $6\times10^{-6}$ after a 10~Myr  evolution\footnote{On our server, the computation of this case, with a time step of $0.08$~day, required about five~days to reach $10$~Myr. This time would, of course, increase with more than one planet in the system and it would probably change on another computer.}.

\subsection{Non-evolving BD: the effect of BD tide}

This case corresponds to Case \textbf{(2)} of Table \ref{tab:sim:tides}, where we switched off the effect of the planetary tide and considered a very dissipative BD.
Initially, the planet is outside the corotation radius, so it migrates outward. 

In agreement with the secular code, the BD tide causes the inclination of the planet to decrease from $\sim12\deg$ to $\sim4.5\deg$ in $10^8$~yr. 
As the planet migrates outward, the rotation period of the BD increases in agreement with the conservation of total angular momentum. 
Indeed, we find that every component of the total angular momentum is conserved. 
In this example, the angular momentum of the BD is of two orders of magnitude higher than the orbital angular momentum of the planet. 
Because $\alpha$ remains below $10^{-4}$ after 100~Myr, we can conclude  that the phenomenon is accurately reproduced here.

\subsection{Non-evolving BD: the  effect of both tides}

This case corresponds to Case \textbf{(3)} of Table \ref{tab:sim:tides}. In this example, the planet is initially outside the corotation radius.

        \begin{figure*}[htbp!]
        \centering
        \includegraphics[width=15cm]{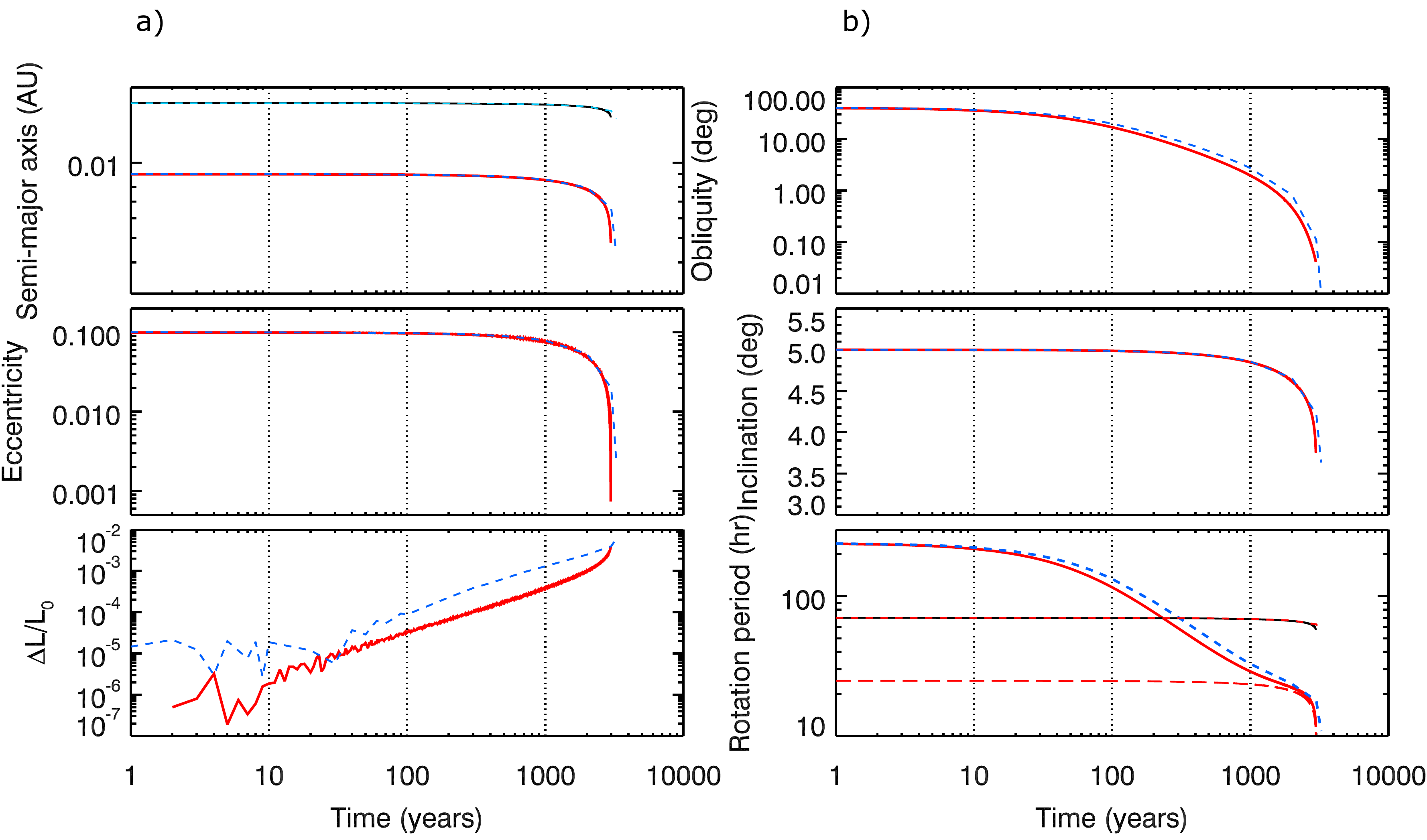}
        \caption{Case \textbf{(3')}: Tidal evolution of a Jupiter-mass planet orbiting a  $0.08~\Msun$ BD, calculated with the secular code (blue dashed line) and the \mTp code (solid red line). Graph a) from top to bottom: evolution of semi-major axis (in red) and of the corotation distance (in black), evolution of eccentricity and conservation of angular momentum. Graph b) from top to bottom: evolution of the obliquity of the planet, evolution of its inclination and evolution of its rotation period (in red) and the BD rotation period (in black). The pseudo-synchronization period is represented in long red dashes.}
        \label{12800_aef_ois}
        \end{figure*}


Figure \ref{12799_aef_ois} shows the evolution of this system. 
We find that the results of \mTp  agree well with those using the secular code. 
The evolution of the different quantities are similar and the quantitative agreement is very good. 
As the planet migrates away, the evolution timescales become longer, which entails a slower evolution, particularly in the late ages of the eccentricity and inclination.

Moreover, the initial heat flux is very strong in comparison to the tidal heat fluxes measured for solar system bodies: $0.08$~W/m$^2$ for Earth \citep{Pollack1993} and between $2.4$ and $4.8$~W/m$^2$ for Io \citep{Spencer2000}. 
Such a large heat flux is likely to have repercussions on the planet's internal structure. 
The high fluxes of Figure \ref{12799_aef_ois}  suggest that the surface and the interior of the planet would melt and that the vertical heat transfer could be very efficient, which does not agree with the dissipation factor value used here. 
As in \citet{Bolmont2013}, we do not include in this work any feedback of the dissipation on the internal structure.

For this system, each components of the total angular momentum $\alpha_i$ is conserved and $\alpha$ reaches a few $10^{-7}$ at a time of $10$~Myr for the \mTp simulation. The total angular momentum here is, therefore, also conserved.

We also test the strength of our code with a more extreme case, for example, Case \textbf{(3')}. 
With an initial orbital distance of $9\times10^{-3}$~AU, the planet is initially inside the corotation radius and thus migrates inward. 
Besides, we consider a BD of a low dissipation factor ($0.01\times\sbd$) so that the planet's BD-tide driven inward migration is not too quick. 

Figure \ref{12800_aef_ois} shows the evolution of this system. 
The planet plunges onto the BD in about $3000$~yr. 
During the inward migration, the eccentricity, obliquity, and inclination all decrease. 
In less than $2000$~yr, the rotation period of the planet evolves from $240$~hr to the pseudo-synchronization period (of about $24$~hr). 
The rotation period of the BD decreases just prior to the fall as a result of  the angular momentum transfer from the planet's orbit to the BD spin. 

Even for this extreme case, both codes lead to the same simulated evolution. 
The collision time may be slightly different, but is of the same order of magnitude for both simulations. 
The bottom-left  panel of Figure \ref{12800_aef_ois} shows the conservation of total angular momentum for this example.

Because the planet ends up  colliding with the BD, we do not expect the conservation of total angular momentum to be perfect. 
Indeed, Figure \ref{12800_aef_ois} shows that for both simulations, $\alpha$ increases with time and reaches about $10^{-2}$ when the collision occurs ($4\times10^{-3}$ for \mTp). 
The planet is initially very close to the BD and gets closer in time, meaning the tidal effects become stronger and stronger. 
This example enables us to test the limits of our model. 
For close-in planets, one should always verify that $\alpha$ is conserved.

In the end, the destiny of the planet is compatible with the theory. 
As its initial orbital distance is less than the corotation distance, the BD tide acts to push the planet inward. 
The qualitative evolution is not likely to change even if the code were to be improved, however the time of collision between the planet and the BD might change.

\subsection{Evolving BD: the effect of both tides}

In Case \textbf{(4)} of Table \ref{tab:sim:tides}, we consider the evolution of the radius and radius of gyration of the BD.

Our code allows us to choose the initial time of the simulations, i.e., the BD age from which we consider the tidal evolution of the planets. 
In this study, we assume that the initial time corresponds to the time of the dispersal of the gas protoplanetary disk \citep[as in][where they discuss the influence of this initial time]{Bolmont2011}. 
We then consider that the planets are fully formed by this time.
The time indicated in the figures corresponds to the time spent after this initial time.

        \begin{figure*}[htbp!]
        \centering
        \includegraphics[width=15cm]{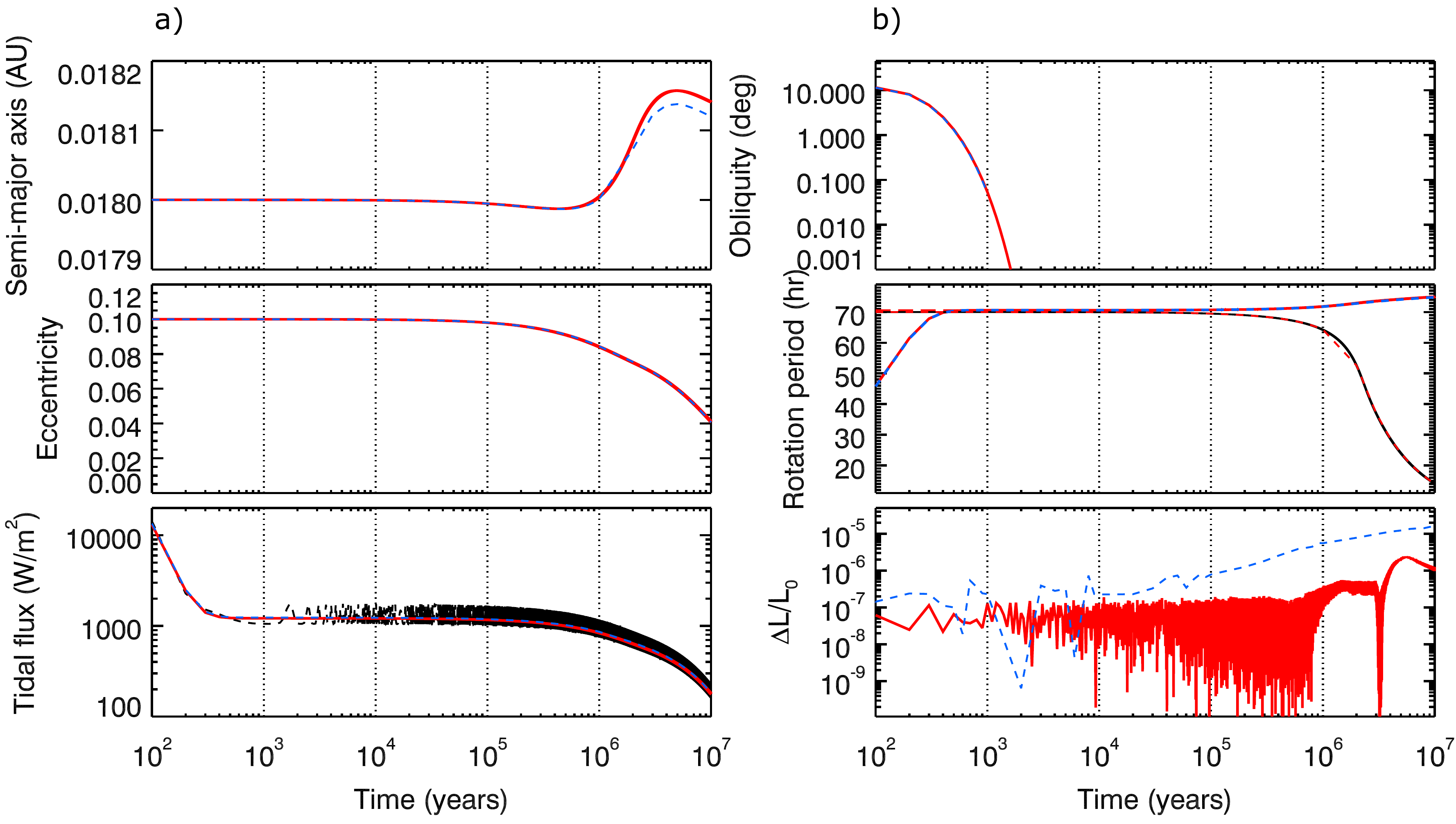}
        \caption{Case \textbf{(4)}: Tidal evolution of an Earth-mass planet orbiting a  $0.08~\Msun$ BD, calculated with the secular code (blue dashed line) and the \mTp code (red line). Graph a) from top to bottom: evolution of semi-major axis (in red) and of the corotation distance (in black), evolution of eccentricity, and evolution of tidal heat flux. Graph b) from top to bottom: evolution of the obliquity of the planet, evolution of its rotation period (in red), the BD rotation period (in black), and the pseudo-synchronization period (red dashed line), and evolution of $\alpha$.}
        \label{12801_aef_osH}
        \end{figure*}

Figure \ref{12801_aef_osH} shows the evolution of this system. 
The evolution calculated with the secular code is in good agreement with the evolution calculated with \mTp.
The competition between the outward migration caused by the BD tide and the inward migration caused by the planetary tide, is well reproduced.

However, the comparison between the outcomes of the two codes shows a difference of $10^{-5}$~AU in the calculated semi-major axis, when the migration direction changes. 
This difference remains small and tends to decrease when the precision of the secular code is increased, which again demonstrates that the  the \mTp code seems more precise than the secular code. 

In any case, the qualitative behavior is reproduced very well, even though small quantitative differences can be seen. 
The \mTp code reproduces  the evolution of the spin of the BD well because of the contraction of its radius (middle panel of  Graph b) in Figure \ref{12801_aef_osH}).
Besides, the total angular momentum is well conserved as can be seen in Figure \ref{12801_aef_osH}.
Indeed, each component of the total angular momentum as well as $\alpha$ remain below $3\times10^{-6}$.

\medskip

These diverse tests show that the tidal integration part of the \mTp code shows a good agreement with the secular code in relation to the orbital evolution of the planet, as well as its rotation state evolution and the rotation evolution of the BD. 
The total angular momentum is always conserved, except when the planet collides with the BD. 
We therefore consider that this code is valid when studying the evolution of tidally evolving multiplanet systems. 
For any simulation, however, one should always make sure that the total angular momentum is conserved.

\subsection{Effect of the rotational induced flattening}\label{rot_flat}

While an Euler integration of the spin may have been sufficient to correctly describe the tidal evolution of the spin of planets, we need to implement a better integrator to accurately describe the precession of the planet's spin axis that results from its own flattening.  
This precession happens on a much shorter timescale than the tidal evolution. 
For the example, in Case \textbf{(5)} the timescales are about a few $10^1$~yr.

Thus, to obtain an accurate integration with \mTp, we need to perform a 5th order Runge-Kutta integration twice in a \mercp time step. 
Dividing the time step into two inside one \mercp time step allows us to be more precise, without demanding too much time. 

We tested the integration of the rotation-induced flattening by comparing our code to the CR13 code. 
In doing so, we obtain similar results for all cases, with a few  quantitative differences.

\begin{figure}[htbp!]
        \begin{center}
        \includegraphics[width=9cm]{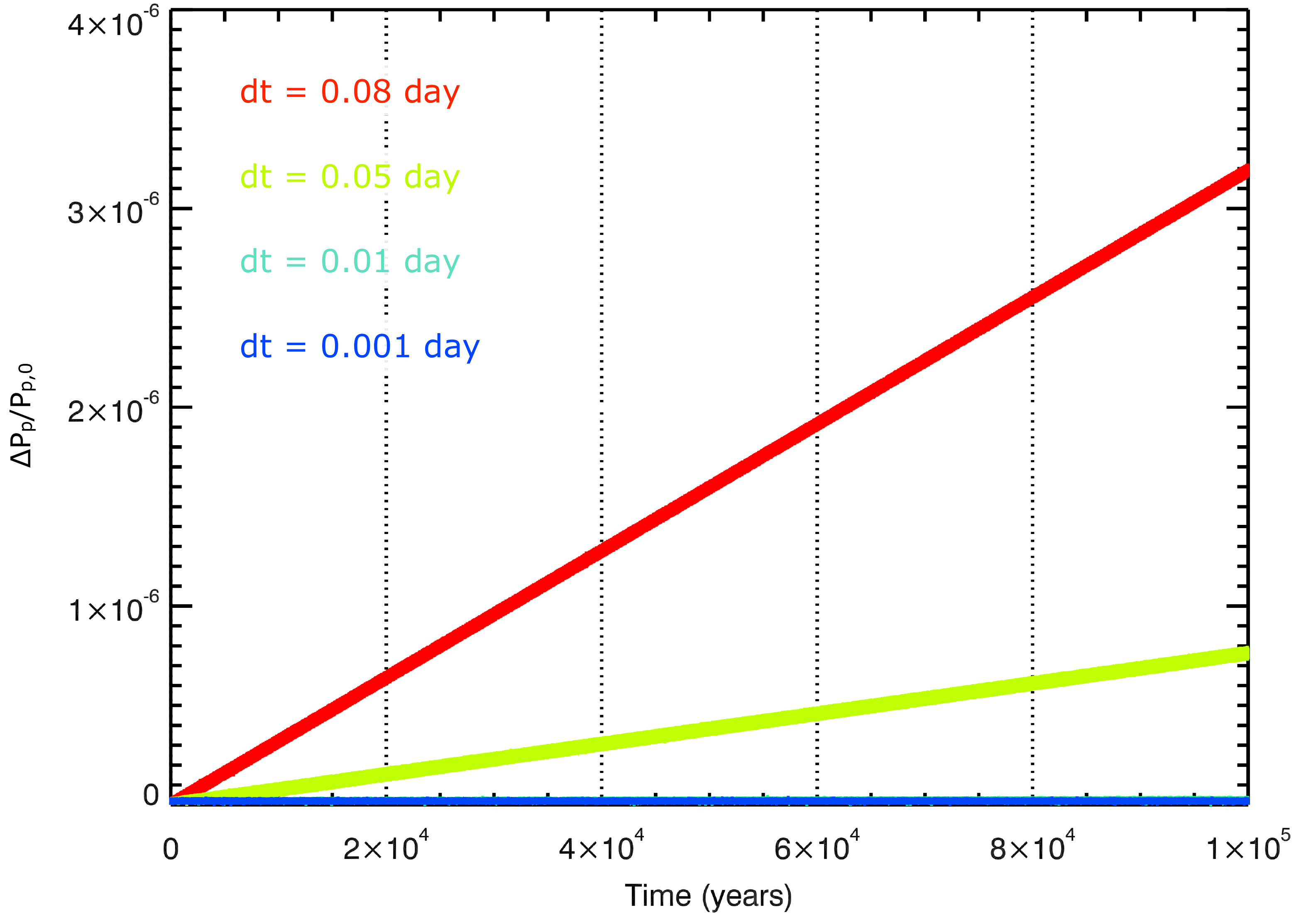}
        \caption{Evolution of $(P_{{\rm p}}-P_{{\rm p},0})/P_{{\rm p},0}$ of the inner planet of the system corresponding to Case \textbf{(6)} of Table \ref{tab:sim:rot_flat}. The colored lines correspond to the results of \mTp: red for a time step of 0.08~day, green for a time step of 0.05~day, light blue for a time step of $0.01,$ and dark blue for a time step of 0.001~day.}
        \label{drift_rot_period}
        \end{center}
        \end{figure}

For  Case \textbf{(6)} of Table \ref{tab:sim:rot_flat}, only the rotational flattening of the inner planet is taken into account.  
The rotation period $P_{{\rm p}}$ (i.e., the norm of the spin) is not influenced by the effect of the rotational flattening. However due to the integration scheme, we observe a small drift in the rotation period of the inner planet (Figure \ref{drift_rot_period}). 
This drift increases linearly with time and decreases when the time step is reduced. 
For a time step of $0.08$~day, i.e., Case \textbf{(6)}, the drift is of about $8\times10^{-5}$~hr after 100~000~yr of evolution, while, for a time step of $0.01$~day, i.e., case \textbf{(6'')}, it is less than $10^{-6}$~hr.
For time steps shorter than $0.01$, the drift is essentially null for  100 000~yr of evolution. 

The shorter the time step, the smaller the differences. From a time step of $0.08$~day to $0.01$~day the improvement is visible, but we can see that there is almost no difference between the light and dark blue curves corresponding to  time steps of $0.01$ and $0.001$~day.
Of course, the execution time is longer for shorter time step, so the time step should be chosen according to the duration of the simulation and the precision needed for the spin of the inner planet of the system.

\begin{figure}[htbp!]
        \begin{center}
        \includegraphics[width=9cm]{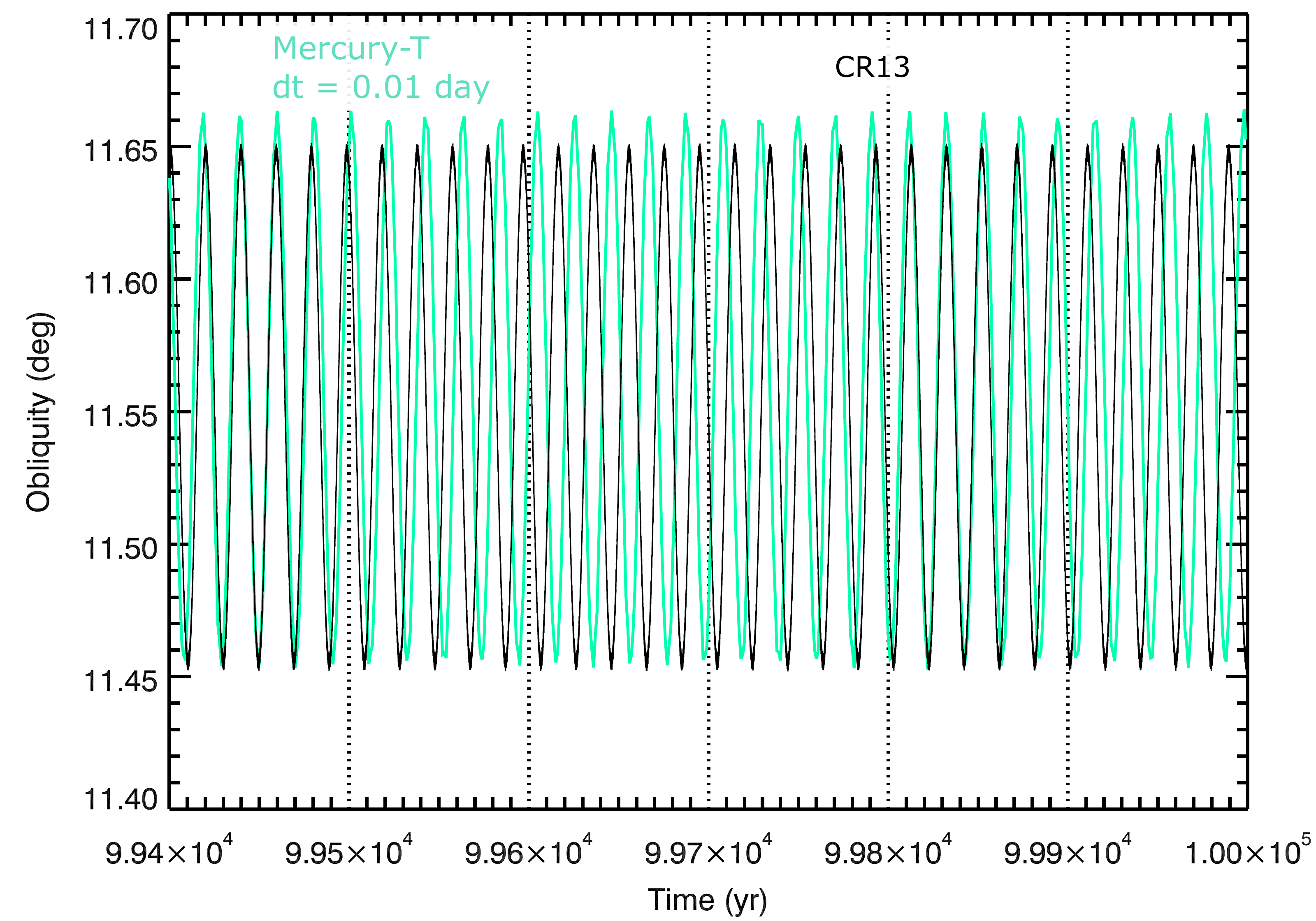}
        \caption{Evolution of the obliquity of the inner planet of the system corresponding to Case \textbf{(6'')} of Table \ref{tab:sim:rot_flat} for the last 600~yr of its evolution. The black line corresponds to the results of the CR13 code. The red line corresponds to the results of \mTp.}
        \label{14973_o_008_001_0001_compAlex}
        \end{center}
        \end{figure}

The semi-major axis of the planets shows perfect agreement, while the eccentricity, the obliquity, the rotation period, and to a lesser extent, the inclination, show a small difference in oscillation frequency. Figure \ref{14973_o_008_001_0001_compAlex} shows the evolution of the obliquity of the inner planet of the system corresponding to Case \textbf{(6'')} of Table \ref{tab:sim:rot_flat} compared with the CR13 code.
The mean value of the obliquity, as well as the maximum and minimum values, are reproduced well. The only difference is the oscillation frequency.

The remaining small difference between \mTp and the CR13 code, i.e., the rotation period drift and the difference in the frequency of the oscillations of quantities, is a result of the different integration schemes and numerical effects.

The system is also very sensitive to the initial conditions.
For example, for a similar rotation period and obliquity, changing the initial direction of the spin of the planet $\mathbf{\Op}$ leads to a different mean value of the oscillations of the obliquity. 

For Case \textbf{(7)}, for which all the effects are considered, the general behavior is perfectly reproduced. 
Both planets migrate outward due to the BD tide and enter a mean motion resonance  at approximately the same time.
By entering the resonance, the eccentricities and obliquities evolve similarly. 

We also tested the conservation of energy and total angular momentum for these examples. 
For Case \textbf{(6)}, we find that each component of the total angular momentum is conserved, and $\alpha$ reaches a few $10^{-7}$ after $10^5$~years of evolution. The total energy of the system is conserved up to a few $10^{-5}$.  
For all the other cases, i.e., \textbf{(6')} to \textbf{(6''')}, the conservation is slightly better, but the orders of magnitude are the same.
In the case of a non-dissipative force, our code conserves total energy and angular momentum.

Apart from very small differences, \textit{Mercury-T} and the CR13 code give the same results. 
We, therefore, consider this agreement good enough for the study of exoplanets.


\section{The case of Kepler-62}

Just as \textit{Mercury-T} can be used to study hypothetical systems, it can also be used for known exoplanet systems. 
This code has been used in the following articles: \citet{Bolmont2013,Quintana2014,Bolmont2014}, and \citet{Heller2014}. 
Here, we present a study of Kepler-62, a system that hosts five planets \citep{Borucki2013}. 
Two of these planets are in the insolation habitable zone (HZ). 

\begin{table}[htbp]
\centering
\caption{Stellar properties} 
\vspace{0.1cm}
\begin{tabular}{ccccc}
\hline     
  Mass  & Radius  & $\ks$ & $\sigma$ & $P_{\star,0}$\\
  ($\Msun$) & ($\Rsun$) & & & (day) \\
\hline
 0.69 & 0.63 & 0.03 & $\ss$ & 79.7 \\
\hline
\end{tabular} 
\label{tab:star} 
\end{table}

\begin{table*}[htbp]
\begin{center}
\caption{Planetary physical parameters}
\vspace{0.1cm}
\begin{tabular}{ccccccc}
\hline
& & Kepler-62b & Kepler-62c & Kepler-62d & Kepler-62e & Kepler-62f \\ 
\hline
Masses ($\Mearth$) &$\mathcal{A}$  &  2.60 & 0.130 & 14 & 6.100 & 3.500 \\
& $\mathcal{B}$    & 2.72 & 0.136 & 14 & 6.324 & 3.648 \\
        a (AU)          &  & 0.0553 & 0.0929 & 0.12 & 0.427 & 0.718 \\
        ecc             &      & 0.071 & 0.187 & 0.095 & 0.13 & 0.094 \\
        inc & & 0.8 & 0.3 & 0.3 & 0.02 & 0.1 \\
$P_{{\rm p},0}$ (hr) && 24 & 20 & 30 & 24 & 24 \\
$\epsilon_{{\rm p},0}$ (rad) && 0.1 & 0.02 & 0.05 & 0.03 & 0.4 \\
\hline
\end{tabular} 
\label{tab:planet} 
\end{center}
\end{table*}

Planetary climate depends on many different parameters, including orbital distance, eccentricity, obliquity, rotation period, and tidal heating \citep{Milankovitch1941,Spiegel2009,Spiegel2010,Dressing2010}. 
Because all of these parameters are influenced by tidal interactions, it is of paramount importance to consider tides in climate studies \citep{Bolmont2014sf2a}. 

Below, we present  a dynamical study of the Kepler-62 system, showing the influence of the diverse physical effects, e.g., tides, rotation flattening, and general relativity, on the stability of the system and on the spin state evolution of the planets. 

To re-create the initial conditions, we used  the data from \citet{Borucki2013} for semi-major axis, eccentricity, longitude of periapsis, and epoch of mid transit. 
We used the values given for the radius of the planets, and we tested the system with different masses for the planets (all of which we assumed to be rocky).

The simulations we show here are, of course, possible evolutions of the system, and we are aware that there are many uncertainties on many parameters, starting with the masses of the planets and their dissipation factors.
However, despite these uncertainties, we aim here to show that some general behaviors can be identified in the dynamics of the system.

\subsection{Dynamics and stabilization}

To investigate the effects of tides, rotation-flattening, and general relativity on the dynamics of the system, we tested the system for five different cases that are listed in Table \ref{tab:effects}.

\begin{table}[htbp]
\centering
\caption{Test simulations for stability}
\vspace{0.1cm}
\begin{tabular}{|c|c|c|c|}
\multicolumn{1}{c}{} &\multicolumn{3}{c}{Effects considered} \\ 
\hhline{~---}
\multicolumn{1}{c}{} &\multicolumn{1}{|c}{GR} &\multicolumn{1}{|c}{Rot. flat.} &\multicolumn{1}{|c|}{Tides} \\
\hhline{~~~~}
\multicolumn{1}{c|}{} &\multicolumn{1}{|c|}{} & &\\
\hline
\textbf{1}   & \xmark & \xmark & \xmark \\
\hline
\textbf{2}   & \cmark & \xmark & \xmark \\
\hline
\textbf{3}   & \cmark & \xmark & \cmark \\
\hline
\textbf{4}   & \cmark & \cmark & \xmark \\
\hline
\textbf{5}   & \cmark & \cmark & \cmark \\
\hline
\end{tabular} 
\label{tab:effects} 
\end{table}

Assuming the planets have a rocky composition (in this hypothesis, planet d has the maximum mass given in \citealt{Borucki2013}), we find that the system, hereafter called system $\mathcal{A}$, is unstable in case \textbf{(1)}. 
After $3$~Myr, planet c is ejected from the system. 
In this case, planet d is very massive: $14~\Mearth$, and its influence on the less massive planet c destabilizes the system. 
We find that system $\mathcal{A}$ is also unstable in case \textbf{(2)}. 
However, the destabilization occurs much later, after $\sim 20$~Myr of evolution. 
Here, the correction for general relativity has  the effect of stabilizing the system. 
General relativity also causes apsidal advance, which can, therefore lead to situations that are favorable or unfavorable  to stability, depending on initial conditions. 
Changing the initial orbital angles, such as the longitude of periastron, modifies the amplitude of the eccentricity oscillations, and this could also lead to a more or less stable system.
However, for our particular choice of initial conditions, it would seem that general relativity has  a stabilizing effect.

Using the same masses for planets b, c, e, and f as in system $\mathcal{A}$, we tested the stability of the system in Cases \textbf{(1)} and \textbf{(2)} for different masses of planet d. 
We found that the system is systematically stable for masses lower than $\sim 8~\Mearth$ for planet~d. 
However for masses higher than $\sim 8~\Mearth$, most simulations lead to destabilization within $30$~Myr. 
For simulations done with a mass higher than $8~\Mearth$ for planet~d, destabilization occurs either for Case \textbf{(1)} or \textbf{(2),} or both, illustrating the importance of taking the correction for general relativity into account.

We also observed that changing the masses of the planet very slightly influences the stability of the system. 
Changing the masses by 5\% leads to a stable system in all cases -- hereafter called system $\mathcal{B}$. 
A broad study of the stability of this system is beyond the scope of this paper, which illustrates  possible future studies using the code \mTp. 
The stability should be tested for all possible masses, which leads a high number of combinations and simulations  to have a map of the stability of the system \citep[e.g.,][]{Laskar1990,Correia2005,Couetdic2010,Mahajan2014}. 
Unstable regions would, therefore, correspond to unrealistic configurations. 
This illustrates the importance of constraining the masses of planetary systems \citep[e.g., with HARPS in][]{Dumusque2014}.

\medskip

When we add tides, as in Case \textbf{(3)}, and assume nominal dissipation factors for the planets, we find that system $\mathcal{A}$ becomes stable for the duration of the simulation. In this case, tides have a stabilizing effect on the system. 
Indeed, in our simulations, both planetary tides and stellar tides, have a damping effect on the eccentricity, therefore reducing the probability of the system to be chaotic. 
System $\mathcal{B}$ remains stable, at least for the duration of the simulation of $30$~Myr. 
Semi-major axes and eccentricities do not significantly  evolve tidally during the simulation, however the obliquities and rotation period of the planets do (see Figure \ref{14376_os_legend}). 

We tested the evolution of system $\mathcal{B}$ for different planetary dissipation factors.
The higher the dissipation of planet j, the faster its tidal evolution.
This effect is first visible on the rotation of the planets (obliquity and rotation period), which evolve much more rapidly.
It also has a small effect on the eccentricity of the planets, which is not visible on the graphs. 
To quantify this, we computed the mean angular momentum deficit \citep[AMD, e.g.][]{Laskar1997} for a set of simulations.

We did a test, varying the dissipation of each planet from a reference simulation corresponding to a dissipation factor of $0.1~\searth$ for all planets.
We increased the dissipation of each planet one after the other from the reference value of $0.1~\searth$ to $10~\searth$ and $100~\searth$.

The results indicate that, by increasing the dissipation of a planet, the AMD decrease. 
The decrease is more or less pronounced depending on the planet considered. 
When increasing the dissipation of planets c, e, and f to $100~\searth$, the AMD is $\sim0.03$\% lower than the reference case.
However, when increasing the dissipation of planet d, the AMD is $0.055$\% lower. 
As planet d is very massive in the system, damping its eccentricity slightly has consequences for the whole system. 
When increasing the dissipation of planet b to $100~\searth$, the AMD is $\sim0.7$\% lower than the reference case. 
Increasing the dissipation of the closest planet has the biggest effect on the dynamics of the system. 
Increasing the dissipation leads to a slightly less chaotic system.

\medskip

When we add the effect of rotation flattening, as in Case \textbf{(4),} we find that system $\mathcal{A}$ is stable, at least for the duration of the simulation of $30$~Myr. This effect stabilizes the system by changing the precession rates. The dynamics of system $\mathcal{B}$ are not significantly changed by this effect.  

\medskip

When we consider all effects , as in case \textbf{(5)}, we find that System $\mathcal{A}$ is stable at least for the duration of the simulation of $30$~Myr. Compared to Case \textbf{(3)}, the addition of the rotation-flattening effect only slightly changes  the equilibrium values of the obliquities of the planets. 

\subsection{Obliquity and rotation period}\label{obl_rot_evolv}

\begin{figure}[htbp!]
        \begin{center}
        \includegraphics[width=9cm]{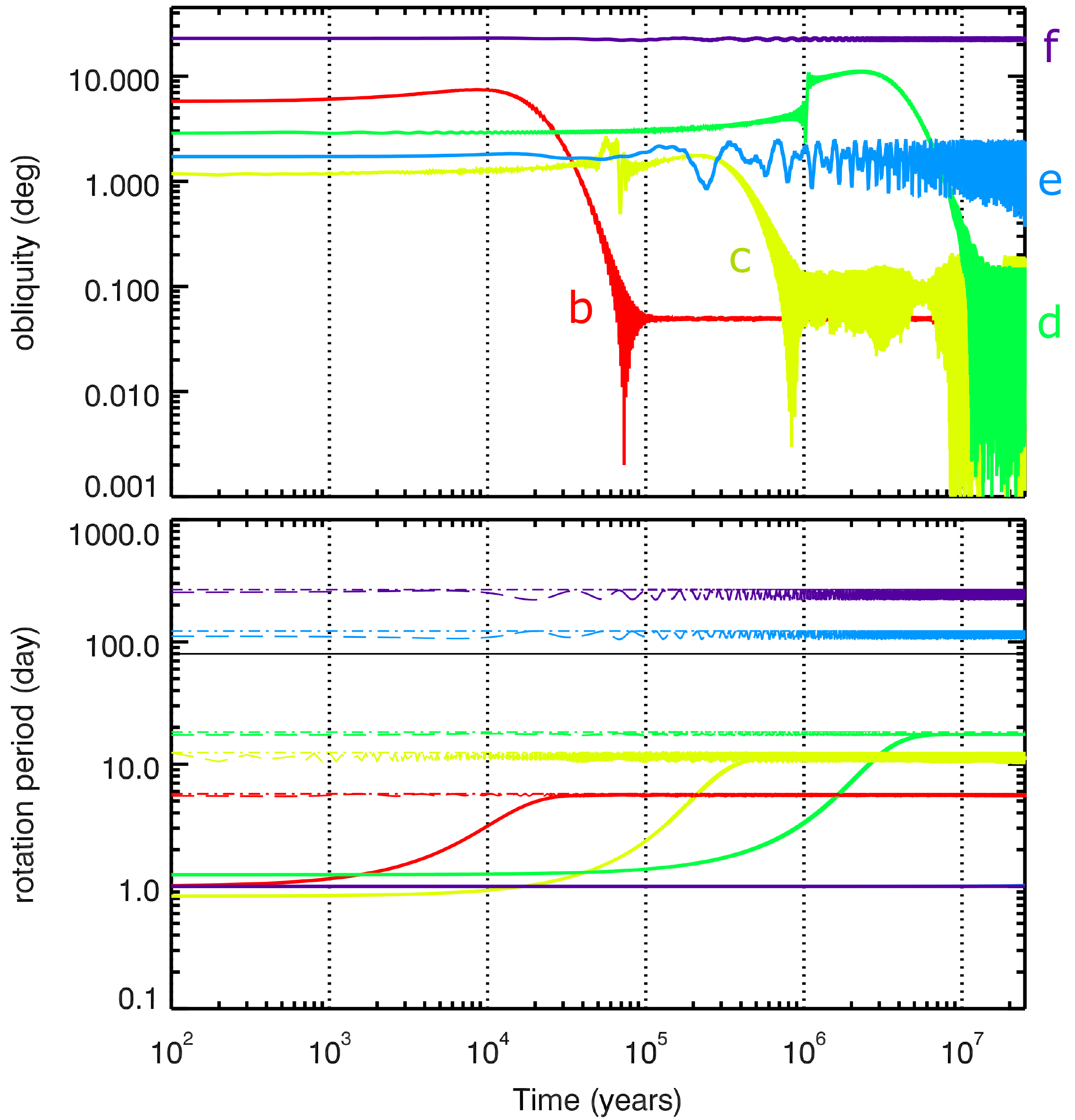}
        \caption{Tidal evolution of the Kepler-62 system ($\mathcal{B}$). Top panel: evolution of the obliquities of the five planets. Bottom panel: evolution of their rotation periods in solid colored  lines. The dashed lines correspond to the pseudo-synchronous rotation period and the solid black line corresponds to the rotation period of the star.}
        \label{14376_os_legend}
        \end{center}
        \end{figure}

Because of the planetary tide, the obliquity decreases and the rotation period evolves toward pseudo-synchronization. 
The evolution timescales of these two quantities are shorter than the timescales of evolution of semi-major axis and eccentricity. Figure \ref{14376_os_legend} shows that, for Kepler-62, the rotation period of the three inner planets of the system evolves toward pseudo-synchronization in less than 10~Myr, and their obliquities evolve toward small equilibrium values ($<1^\circ$). 

These simulations were performed using the values in \citet{Borucki2013} as initial conditions, so that our simulation shows how the system could evolve in the future. 
However, we can draw some conclusions on the past evolution of the system from Figure \ref{14376_os_legend}, or from a simple evolution timescale calculation.
Given that the age of the system is estimated at 7~Gyr \citep{Borucki2013}, we indeed expect that the three inner planets of the Kepler-62 system are now  rotating slowly (their period is higher than 100 hr), and they have quasi null obliquities. 

\begin{figure}[htbp!]
        \begin{center}
        \includegraphics[width=9cm]{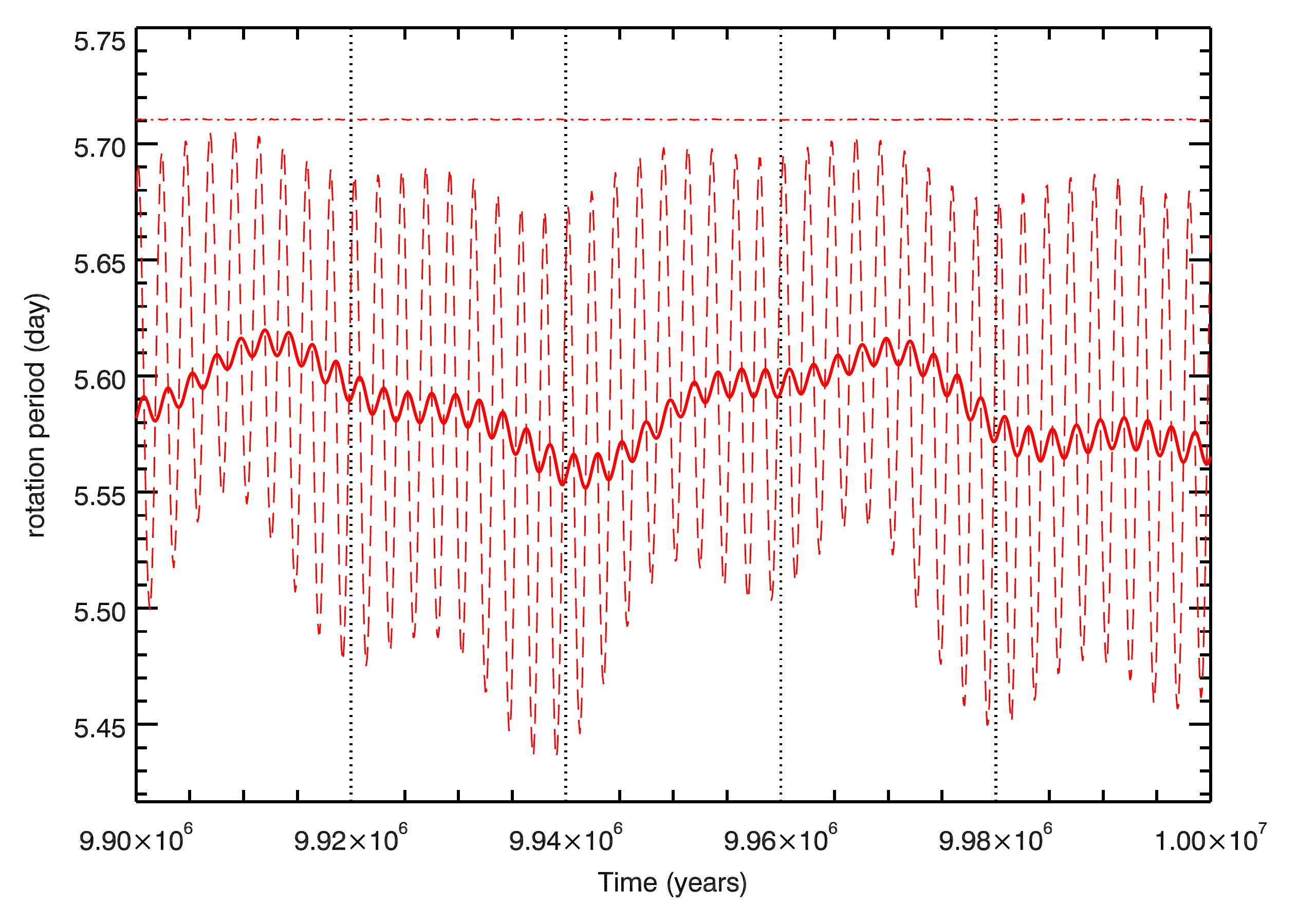}
        \caption{Short-term (100 000-year) evolution of the rotation period of Kepler-62b ($\mathcal{B}$). Solid line: rotation period. Dashed line: pseudo-synchronization period.\ Dashed-dotted line: synchronization period.}
        \label{14376_s}
        \end{center}
        \end{figure}

The ratio between the pseudo-synchronization rate and the orbital frequency  depends only on the eccentricity of the planet. 
But in a multiplanet system, the eccentricity of a planet is excited owing to the planet-planet interactions, and oscillates with a combination of frequencies that correspond to secular modes \citep[e.g.,][]{MurrayDermott1999}. 
In our simulations, the planets experience relatively large eccentricity oscillations (see next section), causing the planets' pseudo-synchronization periods to also oscillate. 
In reality, the rotation periods of the planets are not exactly equal to the corresponding pseudo-synchronization period. 
Figure \ref{14376_s} shows the evolution of the rotation period of Kepler-62b compared to the pseudo-synchronization period and the synchronization period. 
The pseudo-synchronization period oscillates too fast for the rotation period to be able to follow. 
As a consequence, the instantaneous rotation period of Kepler-62b oscillates out of phase with the pseudo-synchronization period and with a lower amplitude.

\begin{figure*}[htbp!]
        \centering
        \includegraphics[width=\linewidth]{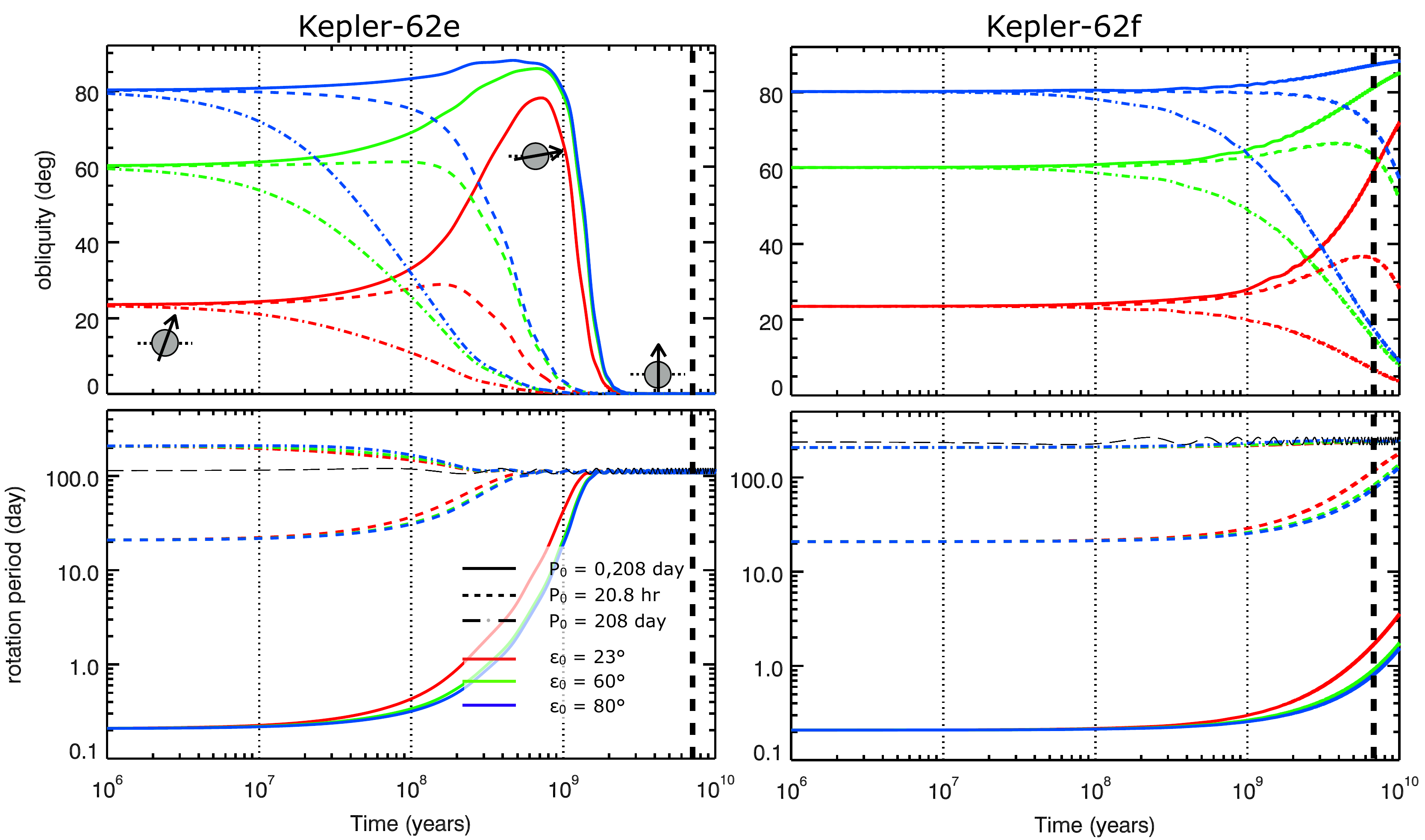}
        \caption{Long-term tidal evolution of the two outer planets of the Kepler-62 system. Top panel: evolution of the obliquities of the five planets. Bottom panel: evolution of their rotation periods in colored solid lines. The dashed lines correspond to the pseudo-synchronous rotation period, and the solid black line corresponds to the corotation radius.}
        \label{Kepler_62_obl_rot}
        \end{figure*}

During the 30 ~Myr of the simulation, the obliquities and rotation periods of the HZ planets Kepler-62e and f did not evolve significantly so we performed longer simulations for these two outer planets. 
Assuming an Earth-like dissipation for the two planets, we found that Kepler-62e is likely today to have reached pseudo-synchronization and have low obliquity. 
Figure \ref{Kepler_62_obl_rot} shows that after 3~Gyr of evolution, the obliquity has been damped and the rotation pseudo-synchronized. 
For Kepler-62f the timescales of evolution are higher and Figure \ref{Kepler_62_obl_rot} shows that the rotation period is still evolving towards pseudo-synchronization after 7~Gyr of evolution, and that the obliquity could still be high. 

The dissipation of the planets is not constrained, and changing the dissipation would only shift the curves right (if the dissipation is lower) or left (if the dissipation is higher). 
As an Earth-like dissipation is probably a high dissipation value \citep[due to the presence of oceans, e.g.][]{Lambeck1977}, it seems likely that the curves should be shifted to the right. 

\subsection{Consequence of dynamics on the potential habitability of Kepler-62e and Kepler-62f}        
        
Kepler-62e and Kepler-62f are both inside the HZ. 
However, being in the HZ does not entail the presence of surface liquid water. 
Surface conditions that are compatible with liquid water depend, of course, not only on the properties of the atmosphere (e.g., pressure, temperature and chemical composition) but also on orbital parameters. 
These include semi-major axis and eccentricity, as well as physical parameters such as the obliquity and the rotation period of the planet \citep{Milankovitch1941}. 

Using the \mTp code, we can simulate the dynamical evolution of habitable planets within their system, taking into account the rich dynamics occurring in a multiplanet system. 
This allows us to provide a set of orbital and physical input parameters that are consistent with the real dynamics
of a planetary system, for any kind of climate model.

        \begin{figure}[htbp!]
        \begin{center}
        \includegraphics[width=9cm]{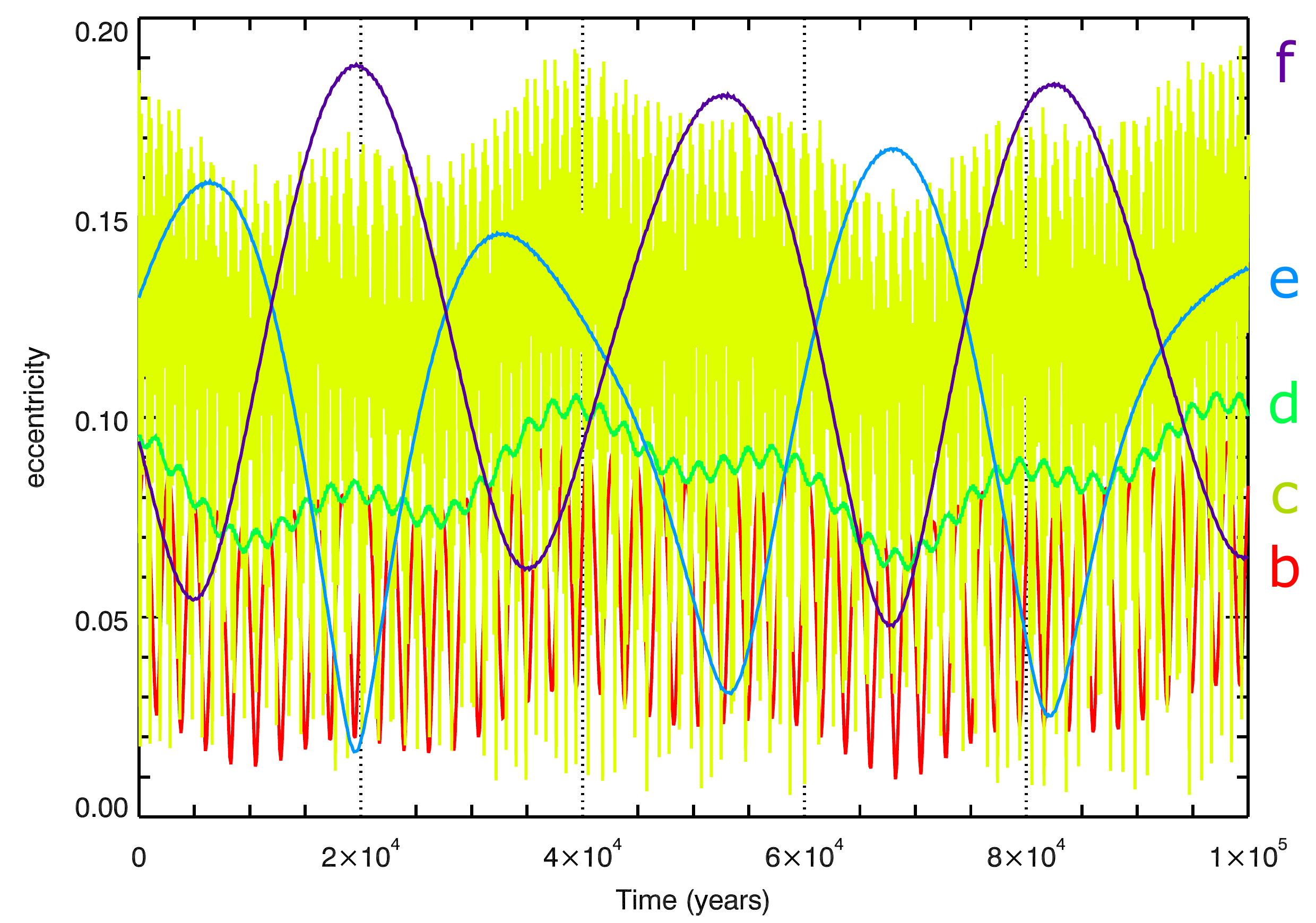}
        \caption{Short-term (100 000~year) evolution of the eccentricity of the Kepler-62 system planets.}
        \label{14376_e_legend}
        \end{center}
        \end{figure}

Figure \ref{14376_e_legend} shows the evolution of the eccentricity of the planets for 100 000~yr. The eccentricities of Kepler-62e and f oscillate respectively between $0.02$ and $0.16$ and between $0.05$ and $0.19$ with a modulated frequency. These important periodical changes in eccentricity have an effect on the climate on Kepler-62e and f, similar to how the Milankovitch cycles had an impact on the paleoclimate of Earth \citep{Berger1992}.

Furthermore, as seen in the previous section, we can also draw some conclusions about the rotation of the planets. 
We have shown that it is likely that Kepler-62e has a pseudo-synchronous rotation (or a near pseudo-synchronous rotation, see section \ref{obl_rot_evolv}) and that its obliquity is very small.
This kind of  planet, with a slow rotation (almost 3 000~hour or 125 day), could have large Hadley cells that  bring hot air to the poles and there would be no longitudinal circulation \citep{MerlisSchneider2010,Leconte2013}. 
However, as Kepler-62e is close to the inner boundary of the HZ, the surface temperatures might not reach low enough values to create cold traps. Consequently, a study of this planet using a global circulation model would be needed to test the potential of this planet to host surface liquid water.

Figure \ref{Kepler_62_obl_rot} shows that Kepler-62f could have a high obliquity and a fast rotation period (as fast as the Earth's 24-hour rotation). 
Of course, we do not know the initial conditions on the spin of the planets. 
However, formation scenarios show that planets are likely to have an initial fast rotation rate because of collisions and  an isotropic distribution of obliquities \citep{KokuboIda2007}. 
For fast rotation rates, the obliquity is excited and can reach high values even if it started at a low value (see, for example, the  solid red line in Figure \ref{Kepler_62_obl_rot}). 
As a result, there is therefore a high probability that the obliquity of Kepler-62f is actually non-negligible. 
Furthermore, its rotation period could still be quite fast: between 20 and 40~hr at the assumed age of the system (see Figure \ref{Kepler_62_obl_rot}).
It is therefore likely that Kepler-62f would have a very different type of climate from its neighbor. 
Indeed, a non-negligible obliquity would lead to seasonal effects, and a fast rotation would lead to a different wind pattern with not only latitudinal winds, but also longitudinal winds.


\section{Conclusions}
\label{Discussion}

In this study, we have presented  a code that computes orbital evolution for tidally evolving multiplanet systems. The theory on which this code is based is the constant time lag model, which is an equilibrium tide model. 
This code allows the user to compute the evolution of the orbital distance, eccentricity and inclination of planets, as well as their rotation state (obliquity and rotation period). It also computes  the rotation period of the host star consistently (taking into account the spin-up due to radius-shrinking and the effects of tides).

The evolution tracks of the radius of various host bodies were implemented: BDs of mass between $0.01$ and $0.08~\Msun$, M dwarfs of $0.1~\Msun$, Sun-like stars and Jupiter. This allows the user to study the influence of a changing radius of the host body on the tidal evolution of planets. 

In this work, we have endeavored  to validate our code. To this end, we compared the outputs of a code that solves the tidal secular equations of single-planet systems \citep[see][]{Bolmont2011,Bolmont2012}, with the outputs of our new code. We also tested the rotational-flattening effect, by comparing \mTp with the CR13 code, which was developed independently.
We found that \mTp reproduces  the secular evolution of the planets well. 
We also made sure that the total angular momentum was conserved in all of our examples. 

Potential users of this code should bear in mind that, when a planet is alone in the system, the code can produce a spurious remnant eccentricity. 
For each simulation, we also advise the user to verify the conservation of total angular momentum and the robustness of the spin integration by doing a simulation without tides and with the effect of the rotational-induced flattening to see whether there is a drift of the mean value of the obliquity. 
If there is a drift, then the time step has to be decreased.

Some ongoing improvements in this code would consist of improving the models of the planets. 
Indeed, the use of the constant time lag model for terrestrial planet has been criticized \citep[e.g.,][]{Makarov2013,Efroimsky2013,MakarovBerghea2013,Correia2014}, and it is probable that the planets are not evolving towards pseudo-synchronization, but are trapped in spin-orbit resonances. 
Besides, to correctly determine the spin of planets, one needs to take thermal tides into account \citep[e.g.][]{Cunha2014,Leconte2015}.
With a global circulation model, \citet{Leconte2015} showed that this phenomenon can drive planets out of synchronization even if they have a thin atmosphere.
We also intend to implement a better description of the dissipation within the star \citep[e.g., using models found in][]{Auclair-Desrotour2014}. 
A wind prescription will also be  added soon (as in Bolmont et al. 2012). 
In the future, we also intend to investigate  the multibulge effect, i.e., the influence of the bulge raised on the star by planet j on the dynamical evolution of planet i$\neq$j \citep[as was achieved in][for the Earth-Moon-Sun system]{ToumaWisdom1994}.

\mTp is a very powerful tool for simulating the evolution of any kind of planetary system. 
It can be used to simulate known exoplanetary systems to try to: identify trends, as we have done in this work for the Kepler-62 system; investigate the stability of the system, taking all the important physical phenomena into account; and to investigate the influence of tidal dissipation factors on the evolution of the system, to maybe constrain the parameters space.
For example, \citet{Bolmont2013} used a previous version of the code to evaluate the possible eccentricity of the transiting inner planet of the 55 Cancri system. 
Using  this code, they investigated if tidal heating could contribute significantly to 55 Cancri e's thermal emission. 
Our code also allows the user to have an idea of the spin state of planets \citep[as in this work or in][which focuses on the Kepler-186 system {and} also constitutes a fine example of the use of \mTp]{Bolmont2014}.

\mTp is particularly interesting to use for simulating the orbital dynamical evolution of habitable planets because it allows for reasonable and consistent input in climate models to investigate the potential of these planets to host surface liquid water and also to investigate the influence of eccentricity oscillations on such a climate.


\begin{acknowledgements}  
The authors would like to thank John Chambers for his  \mercp \ code and Christophe Cossou for having updated the \mercp code in fortran 90, and for his help in developing this code. 

E. B. acknowledges that this work is part of the F.R.S.-FNRS ``ExtraOrDynHa'' research project.

A.C. acknowledges support from CIDMA strategic project UID/MAT/04106/2013.

The authors would like to thank Rosemary Mardling for a thorough referee report that helped them improve the quality of the manuscript.

\end{acknowledgements}


\end{document}